\documentclass[
superscriptaddress,
 amsmath,amssymb,
 aps,
prstab,
twocolumn,
]{revtex4-2}

\usepackage{graphicx}
\usepackage{dcolumn}
\usepackage{bm}
\usepackage{paralist}
\usepackage{epstopdf}
\usepackage{epsfig}
\usepackage{natbib}
\usepackage{textcase}
\usepackage{bm}
\usepackage{url}
\usepackage{makecell}
\usepackage{tabularx}
\usepackage{multirow}
\usepackage{array}
\usepackage{wrapfig}
\usepackage{float}

\begin{document}


\title{Hot Isostatic Pressing (HIP) assisted diffusion bonding between CuCr1Zr and AISI 316L for application to the Super Proton Synchrotron (SPS) internal beam dump at CERN}
\author{S.~Pianese}
\email[] {stefano.pianese@cern.ch}
\author{A.~Perillo Marcone}
\email[]{antonio.perillo-marcone@cern.ch}
\author{F.-X.~Nuiry}
\author{M.~Calviani}
\email[]{marco.calviani@cern.ch}
\author{K.~Adam Szczurek}
\author{G.~Arnau Izquierdo}
\author{P.~Avigni}
\author{S.~Bonnin}
\author{J.~Busom Descarrega}
\author{T.~Feniet}
\author{K.~Kershaw}
\author{J.~Lendaro}
\author{A.~Perez Fontenla}
\affiliation{CERN, 1211 Geneva 23, Switzerland}
\author{T.~Schubert}
\affiliation{IFAM, Winterbergstrasse 28, 01277 Dresden, Germany}
\author{S.~Sgobba}
\affiliation{CERN, 1211 Geneva 23, Switzerland}
\author{T.~Wei{\ss}g{\"a}rber}
\affiliation{IFAM, Winterbergstrasse 28, 01277 Dresden, Germany}






\date{\today}

\begin{abstract}
The new generation internal beam dump of the Super Proton Synchrotron (SPS) at CERN will have to dissipate approximately 270~kW of thermal power, deposited by the primary proton beam. For this purpose, it is essential that the cooling system features a very efficient heat evacuation. Diffusion bonding assisted by Hot Isostatic Pressing (HIP) was identified as a promising method of joining the cooling circuits and the materials of the dump’s core in order to maximise the heat transfer efficiency. This paper presents the investigation of HIP assisted diffusion bonding between two CuCr1Zr blanks enclosing SS 316L tubes and the realisation of a real size prototype of one of the dump’s cooling plate, as well as the assessments of its cooling performance under the dump’s most critical operational scenarios. Energy-dispersive X-ray (EDX) spectroscopy, microstructural analyses, measurements of thermal conductivity and mechanical strength were performed to characterize the HIP diffusion bonded interfaces (CuCr1Zr-CuCr1Zr and CuCr1Zr-SS~316L). A test bench allowed to assess the cooling performance of the real size prototype. At the bonded interface, the presence of typical diffusional phenomena was observed. Moreover, measured tensile strength and thermal conductivity were at least equivalent to the lowest ones of the materials assembled and comparable to its bulk properties, meaning that a good bonding quality was achieved. Finally, the real size prototype was successfully tested with an ad-hoc thermal test bench and with the highest operational thermal power expected in the new generation SPS internal beam dump. These results demonstrated the possibility to use HIP as manufacturing technique for the cooling plates of the new generation SPS internal beam dump, but they also open up the way for further investigations on its exploitability to improve the cooling performance of any future high intensity beam intercepting device or in general devices requiring very efficient heat evacuation systems.

\end{abstract}

\maketitle


\section{Introduction}
\begin{figure}
    \centering
    \includegraphics[width=0.5\textwidth]{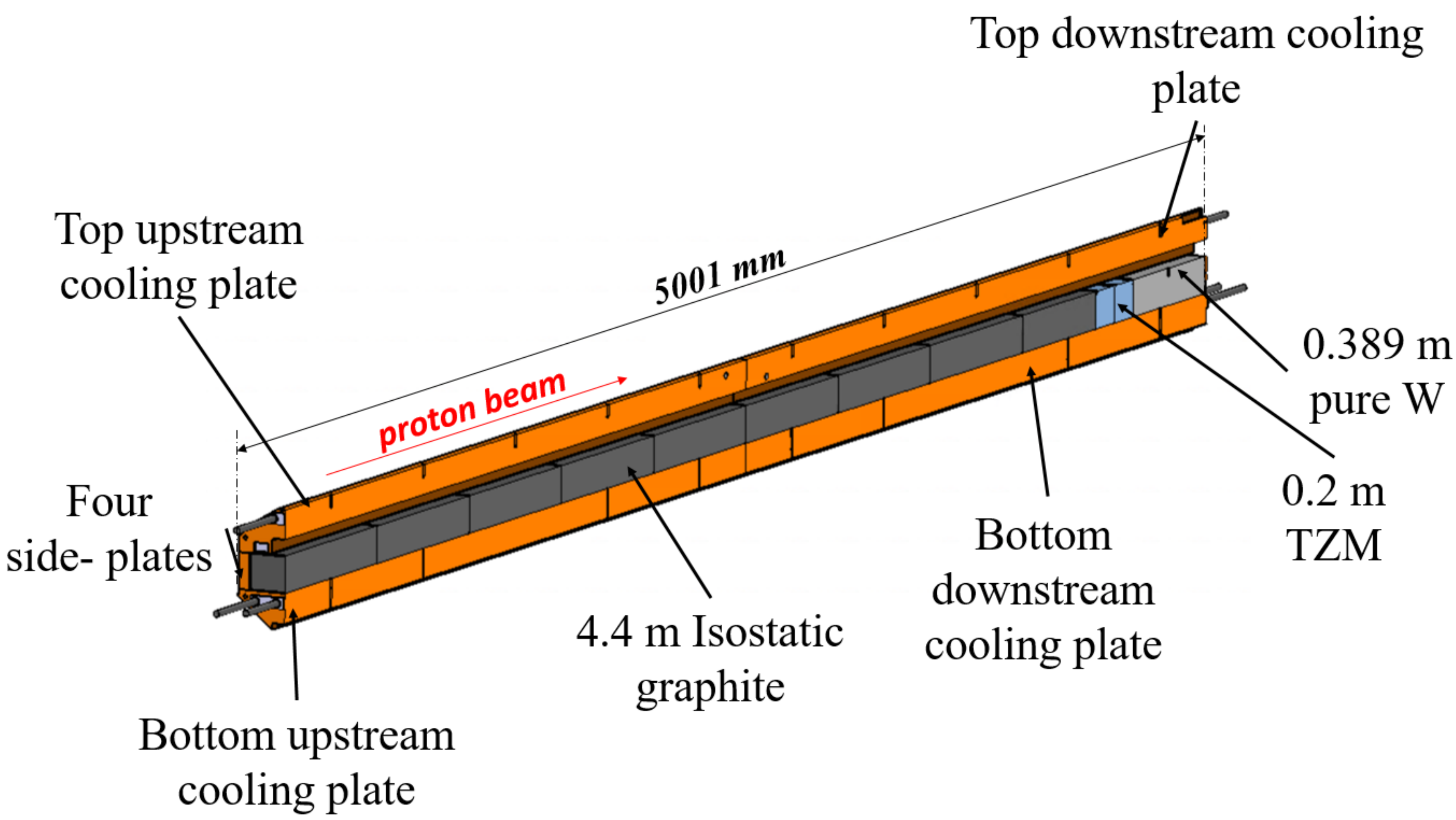}
    \caption{Cross section of the new generation SPS internal dump core, TIDVG\#5}
    \label{fig:CrossSectionTIDVG}
\end{figure}

The new generation internal beam dump (Target Internal Dump Vertical Graphite, TIDVG\#5~\cite{TIDVG5} of the Super Proton Synchrotron (SPS) at CERN, built in the framework of the Large Hadron Collider (LHC) Injectors Upgrade (LIU) Project~\cite{LHC_Upgrade} has been installed in the long straight section 5 (LSS5) of the SPS during the CERN’s Long Shutdown 2 (LS2), 2019-2020. Higher beam intensities and energies, resulting from the LIU upgrades, are making the cooling of this device, and in general of most beam intercepting devices at CERN, increasingly demanding. A key problem is obtaining high levels of heat transfer efficiency between the core beam absorbing materials and the cooling circuits. Specifically, the SPS internal beam dump is required to evacuate a maximum average beam thermal power of approximately 270~kW, more than 4 times the limit of its predecessor TIDVG\#4~\cite{TIDVG4}. As shown in Figure~\ref{fig:CrossSectionTIDVG}, the core of the SPS internal dump is made of an array of absorbing blocks consisting in 4.4 m of isostatic graphite, 0.2 m of TZM and $\sim$0.4 m of pure tungsten, enclosed within 2.5 m long CuCr1Zr cooling plates and under ultra-high vacuum conditions (UHV). Whenever high energy proton beams need to be dumped (i.e. in case of emergency, during LHC beam setup or LHC filling, machine developments and to dispose of the part of the beam for fixed targets remaining after the slow-extraction process), they are deflected onto the absorbing blocks by an upstream set of three vertical and three horizontal kickers. The CuCr1Zr cooling plates must dissipate most of the thermal power carried by the proton beam and diffused by conduction from the absorbing blocks to the CuCr1Zr heat sinks. As a consequence, their heat evacuation efficiency is crucial for the survivability of the absorber system. 

In the previous generation device~\cite{Ross:702618}, the CuCr1Zr plates were mechanically clamping stainless steel tubes. However, interfaces play a major role in the heat evacuation mechanism as they represent a resistance to the heat transfer. For this reason, a possible way to obtain the highest possible thermal contact conductance was identified by means of diffusion bonding. Hot Isostatic Pressing assisted diffusion bonding was previously identified as promising joining technique between copper alloys and stainless steel~\cite{HIPing_CuSS}; by combining elevated temperatures (up to 2000°C) and isostatic gas pressure (up to 200~MPa), HIP is mainly used to heal casting or additive manufacturing defects and for powder metallurgy parts consolidations~\cite{HIP_Overview,HIP_AddMan,TI}. Although HIP has been extensively studied as joining technique for several materials~\cite{HIP_Refractory,ITER,MuonTarget}, it has not been used so far to manufacture by diffusion bonding large cooling plates for ultra-high vacuum (UHV) applications. The core of the new generation SPS beam dump is 5 m long, but, given the maximum dimensions of HIP ovens available, the HIPed cooling plates were divided in four parts of 2.5 m each: two bottom upstream/downstream and two top upstream/downstream (Figure~\ref{fig:CrossSectionTIDVG}). The objective of this work was to characterize the HIP diffusion-bonded interfaces between CuCr1Zr and SS~316L, as well as between CuCr1Zr and CuCr1Zr, and to assess the cooling performance of a real size prototype of cooling plate under the dump’s most critical operational scenario.  

The use of HIPing to diffusion bond CuCr1Zr with SS316L has also been used with success for the CERN Proton Synchrotron (PS) Internal dump~\cite{PS_Dump} as well as for the Beam Dump Facility target prototype beam test~\cite{BDF} to diffusion bond dissimilar refractory metals. Because of increasingly demanding particle beams, this manufacturing technique could be of interest to improve the cooling performances of many other beam intercepting devices (BIDs) such as the new ISOLDE beam dumps, planned to be refurbished by 2025 within the scope of the EPIC Project~\cite{ISOLDE}, the new 30 kW beam dump of the J-PARC Main Ring (MR)~\cite{J_Parc}, as well as absorbing devices within the Proton Improvement Plan-II (PIP-II) Project~\cite{PIP} at FNAL, the International Linear Collider (ILC) photon dumps or in general any other device requiring very efficient heat evacuation systems.

\section{Methodology}
The needed components for a real size cooling plate prototype were manufactured, assembled at CERN and diffusion bonded by a HIP cycle. A sub-part of the prototype was cut and used to manufacture samples to characterize the diffusion bonded interfaces by means of microstructural analyses, Energy-dispersive X-ray (EDX) spectroscopy and measurements of thermal conductivity and mechanical strength. Measurements of electrical conductivity and Brinell hardness were performed on the bulk of CuCr1Zr to quantify the evolution of these properties after the HIP cycle and the following thermal treatments. 
The rest of the prototype was used within a built-in-house test bench to assess its cooling performance under the dump’s most critical operational scenario.

\subsection{Materials}
The prototype to be assembled by HIP diffusion bonding (Figure~\ref{fig:ProtoDesign}) consisted of two precipitation hardened CuCr1Zr blanks from ZOLLERN GMBH (Germany), 2.7 m long, embedding in a sandwich two U-bent SS~316L (EN 1.4435) tubes. Table~\ref{table1} and Table~\ref{table2} show the chemical composition of this copper alloy and the SS~316L, respectively. The SS 316L cooling tubes are cold bent and annealed. The annealing cycle for SS~316L consists in a heating-up phase at 200-300~°C/h up to 950~°C, followed by a plateau at this temperature of at least 2h and natural cooling-down to ambient temperature. The furnace employed was of the type “Grand-Four” UGINE for thermal treatments under vacuum, manufactured by Alstom in 1974. It has an exploitable inner diameter of 990~mm for 6000~mm height and a maximum capacity of 1000~kg. The maximum reachable temperature is 1100~°C with heating rates ranging from 50~°C/h to 300~°C/h.  The cooling is for natural convection. 
The CuCr1Zr blanks are multi-directionally forged according to the standard EN~12420 and heat treated as described below. Refs.~\cite{TTH_Cu,Recovery_Cu} indicate that the following thermal treatments sequence, applied on CuCr1Zr, allows to obtain the highest values of electrical and thermal conductivity, hardness, tensile properties, good ductility and machinability:
\begin{itemize}
    \item Solution annealing between 950-1000~°C for 30~min followed by water quenching;
\item Precipitation hardening through ageing at 480-500~°C for 2~h and slow cooling in air.
\end{itemize}
Hereinafter in the paper, the material shall be considered to have followed this treatment and it will be referred as “precipitation hardened CuCr1Zr” or simply “CuCr1Zr”.

After the HIP cycle, despite the high temperatures, the grains’ structure does not change, most likely due to the hydrostatic pressure at which the material is exposed during treatment, but the thermal and mechanical properties need to be recovered using the same treatment. Table~\ref{table3} shows how electrical conductivity and Brinell hardness, measured on the rear side of the prototype, change after HIP and following the different steps of the heat treatment cycle. For this application, thermal conductivity was slightly more important than mechanical strength, therefore the aging phase was performed closer to the upper limit of 500 °C. Indeed, a higher temperature promotes the precipitation but induces also a slight coarsening of the precipitates thus lowering the mechanical strength. 

\begin{table*}[h!tbp]
\centering
\caption{CuCr1Zr chemical composition [wt.\%] and material designation according to EN 12420 and product analysis} 
\begin{tabular}{|m{1.5cm}|m{2.5cm}|m{1.5cm}|m{1.5cm}|m{0.6cm}|m{0.5cm}|m{0.6cm}|m{0.6cm}|m{0.7cm}|m{1cm}|}
 \hline
 \multicolumn{2}{|c|}{Material designation} & & Element & Cu & Cr & Fe & Si & Zr & Others total\\
 \hline
 Symbol & Number & EN12420 & Min/Max & Rem - & 0.5/ 1.2 & -/ 0.8 & -/ 0.1 & 0.03/ 0.3 & -/ 0.2\\
\hline
CuCr1Zr & EN: CW106C; UNS: C18150 & \multicolumn{2}{|c|}{Analysis product}& Rem & 0.8 & $<$.01 & $<$.01 & .09 & -\\
\hline
\end{tabular}
\label{table1}
\end{table*}

\begin{table*}[h!tbp]
    \centering
    \caption{SS316L chemical composition [wt.\%] and material designation according to EN10088-3 and product analysis}
    \begin{tabular}{|m{1.6cm}|m{1.2cm}|m{0.7cm}|m{0.7cm}|m{0.7cm}|m{0.7cm}|m{0.7cm}|m{0.7cm}|m{0.7cm}|m{0.7cm}|m{0.7cm}|m{0.7cm}|}
    \hline
         Material designation& & C & Si & Mn & P & S & N & Cr & Cu & Mo & Ni \\
         \hline
         \multirow{2}{4em}{AISI 316L} & EN: 10088-3 & 0.030 & 1.00 & 2.00 & 0.045 & 0.030 & 0.11 & 17.0/ 19.0 & - & 2.50/ 3.00 & 12.5/ 15.0\\
         \cline{2-12}
         & Product analysis & 0.024 & 0.42 & 1.54 & 0.029 & 0.008 & 0.045 & 17.59 & - & 2.56 & 13.08 \\
         \hline
    \end{tabular}
    \label{table2}
\end{table*}

\begin{table*}[h!tbp]
    \centering
    \caption{Electrical conductivity and Brinell hardness evolution with heat treatments after HIP}
    \begin{tabular}{|m{4cm}|m{1.8cm}|m{1.5cm}|}
    \hline
         Heat treatment & Electrical conductivity & Hardness \\
         \hline
         HIP & 41.7 MS/m & 45 HB \\
         \hline
         Solution annealing and water-quenching & 25 Ms/m & 60 HB \\
         \hline
         Ageing & 51 MS/m & 108 HB\\
         \hline
    \end{tabular}
    \label{table3}
\end{table*}

\subsection{Prototype design}
\begin{figure}
    \centering
    \includegraphics[width=0.5\textwidth]{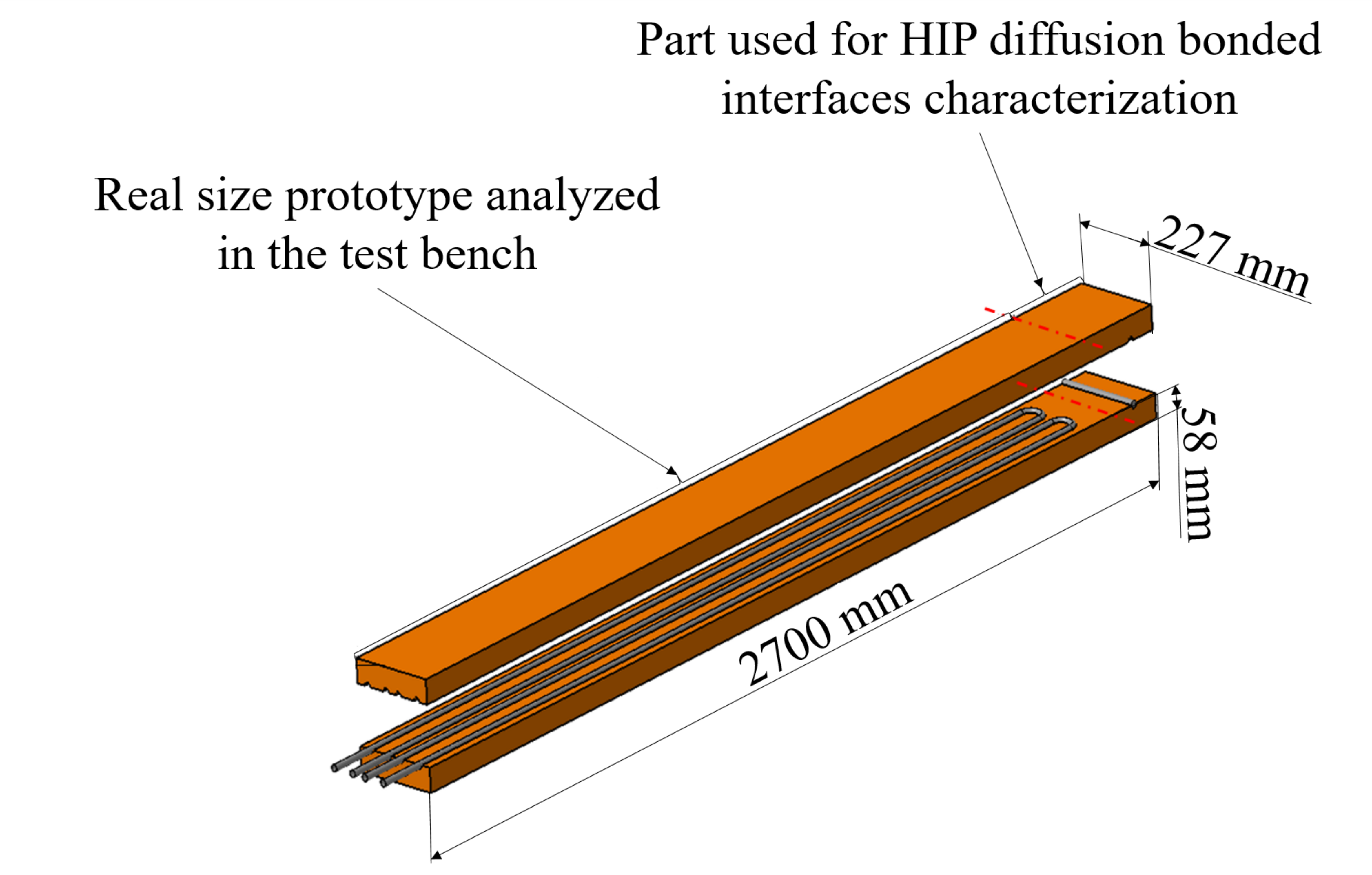}
    \caption{Design details of the real size cooling plate prototype.}
    \label{fig:ProtoDesign}
\end{figure}

Figure \ref{fig:ProtoDesign} shows the overall dimensions of the two precipitation hardened CuCr1Zr halves assembled with the tubes, corresponding to 227 mm (W) × 116 mm (H) × 2700 mm (L).

The tubes inner and outer diameters are 15 mm and 18 mm, respectively.
The rear part of the assembly, containing a straight tube of SS~316L, is meant to be cut off to be used for the characterization of the bonded interfaces after the HIP cycle and heat treatments. Before the HIPing cycle, to ease as much as possible the diffusion bonding at the interfaces, the parts have been cleaned. The cleaning consists in de-greasing the parts by immersion in a solution of a compatible alkaline detergent (pH 9.7) at 60°C for 60 minutes, while an ultrasonic agitation is set for 10 to 20 minutes, and final rinsing with demineralized water followed by drying with clean compressed air or nitrogen.

\subsection{HIP capsule}
\begin{figure}
    \centering
    \includegraphics[width=0.5\textwidth]{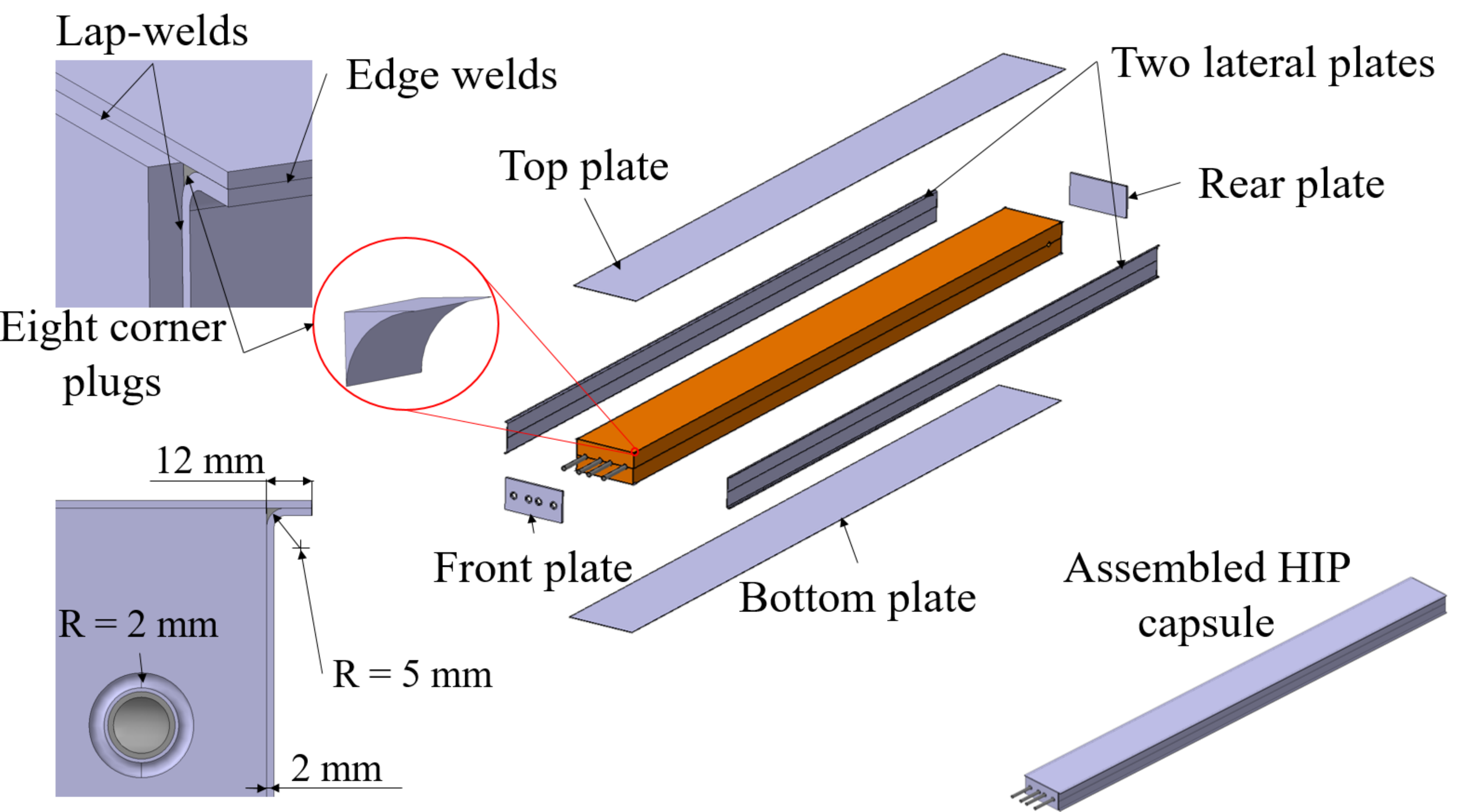}
    \caption{HIP capsule components and main characteristics.}
    \label{fig:HIPCapsule}
\end{figure}

During the HIP cycle, pressure is hydrostatically applied inside the tubes and outside the precipitation hardened CuCr1Zr blanks. In order to promote diffusion bonding at the components’ interfaces, these latter must be kept under vacuum during the HIPing cycle. As shown in Figure~\ref{fig:HIPCapsule}, a capsule encloses the assembly and keeps the vacuum leak tightness all over the cycle duration. The HIP capsule is made of six SS~304L plates: the front and rear plates are 8 mm thick whereas the rest is 2 mm. The plates are welded together, the lateral and top/bottom by edge welds, while the front and rear by lap welds with the rest of the plates. In order to lower as much as possible the stresses on the welds during the HIP cycle, the capsule dimensions are tailored on the inner CuCr1Zr/tubes assembly. The eight corners formed by the six SS~304L plates are filled with “plugs”, which are melted during the welding process making more uniform beads. All the welds of the capsule are tested by dye penetrant and total helium leak test. The dye penetrant tests are performed according to the standard ISO 3452-1-2013 and all the welds should be conform at least to the acceptance level 2X of the standard ISO 23277-2009-EN-BS. The penetrant liquid, the developer and cleaner used are respectively Ardrox penetrant 907 PB (red dye), Ardrox 9D1 B and Ardrox 9 PR 88. The total helium leak test is performed according to the standard EN 1779 and NF EN ISO 20485. The capsule is evacuated and connected to a detector type Leybold L300 or ASM 142 calibrated on helium-4 ( the gas tracer). A higher concentration of the gas tracer is created around the capsule by enclosing it in a bag. The leak rate is calculated based on the signal recorded and the helium-4 concentration inside the bag. The capsule is accepted if there are no leaks within the sensibility of the test, $1 E^{-9} mbar.l/s$.    

\begin{figure}
    \centering
    \includegraphics[width=0.5\textwidth]{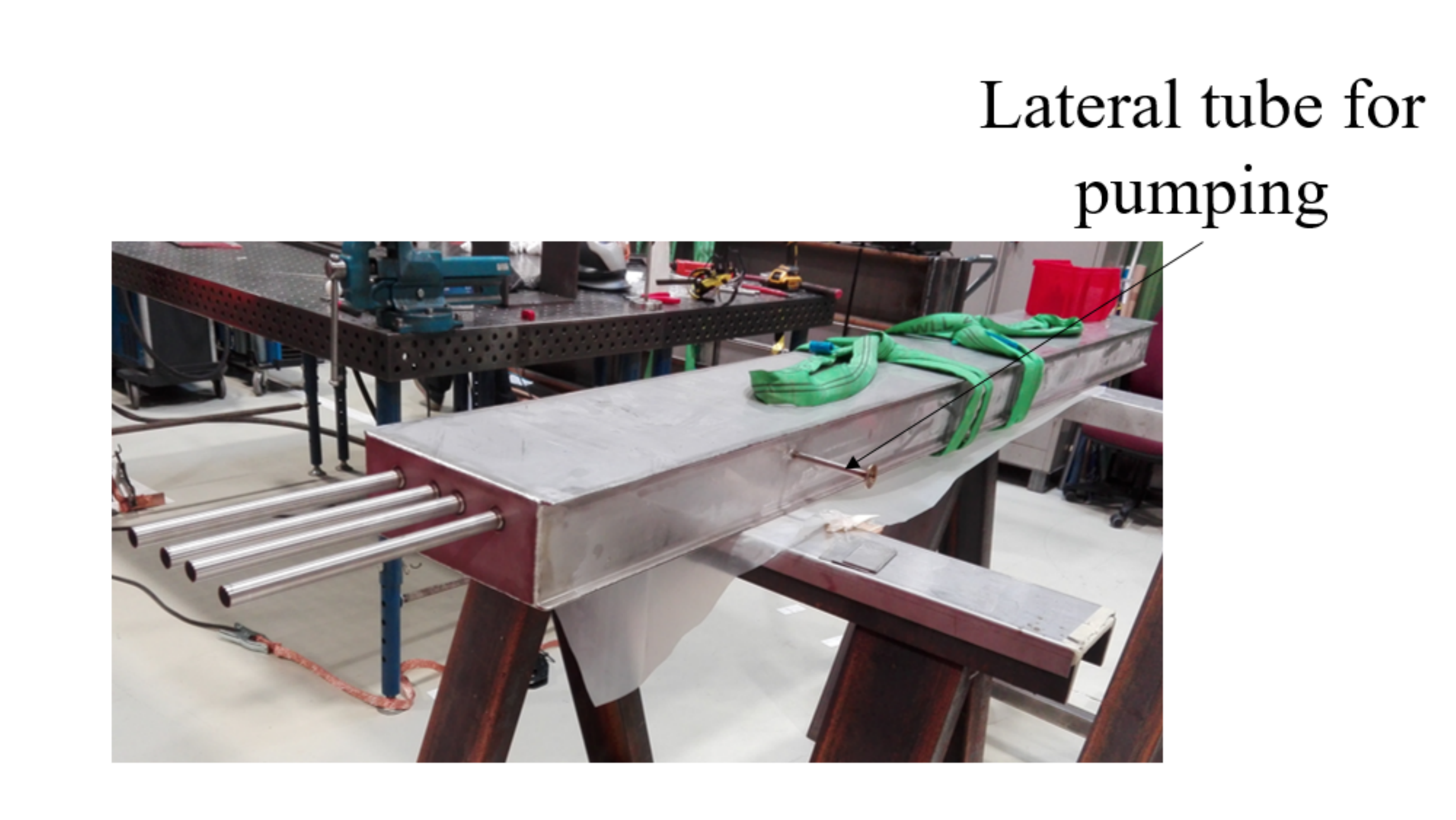}
    \caption{The HIP capsule containing the CuCr1Zr and tubes assembly, PT and He-leak tested with the lateral tube for pumping}
    \label{fig:HIPCapsule2}
\end{figure}

It is important to reach a good vacuum level (at least $10^{-3} mbar$) in order to avoid that air pockets remain trapped at the interfaces thus potentially lowering the bonding quality. Before the HIP cycle, the capsule is pumped down by means of the lateral tube shown in Figure~\ref{fig:HIPCapsule2}. CuCr1Zr, SS~304L and SS~316L have very similar thermal expansion coefficients within the operational HIP temperatures, ranging from ~ $15*10^{-6} m/m K^{-1}$ at room temperature up to almost $20*10^{-6} m/m K^{-1}$ at 1000~°C: stresses due to relative thermal expansions and contractions are then limited.

\subsection{HIPing cycle}
The capsule is transported into a HIP unit: a system applying high pressure with argon gas and high temperature. Table~\ref{table4} shows the main characteristics of the HIP unit used.

\begin{table*}[h!tbp]
    \centering
    \caption{HIP unit characteristics}
    \begin{tabular}{|m{2cm}|m{1.8cm}|m{1.3cm}|m{1.5cm}|m{1.5cm}|m{1.5cm}|}
    \hline
         Manufacturer & Model & Max load dimensions [mm]& Max load weight [kg] & Max Temperature [°C] & Max pressure [MPa] \\
         \hline
         ASEA & QIH307 "Mega-HIP" & $\phi$ ~ 1300 h ~ 3200 & 10000 & 1240 & 140\\
         \hline
    \end{tabular}
    \label{table4}
\end{table*}

\begin{figure}
    \centering
    \includegraphics[width=0.5\textwidth]{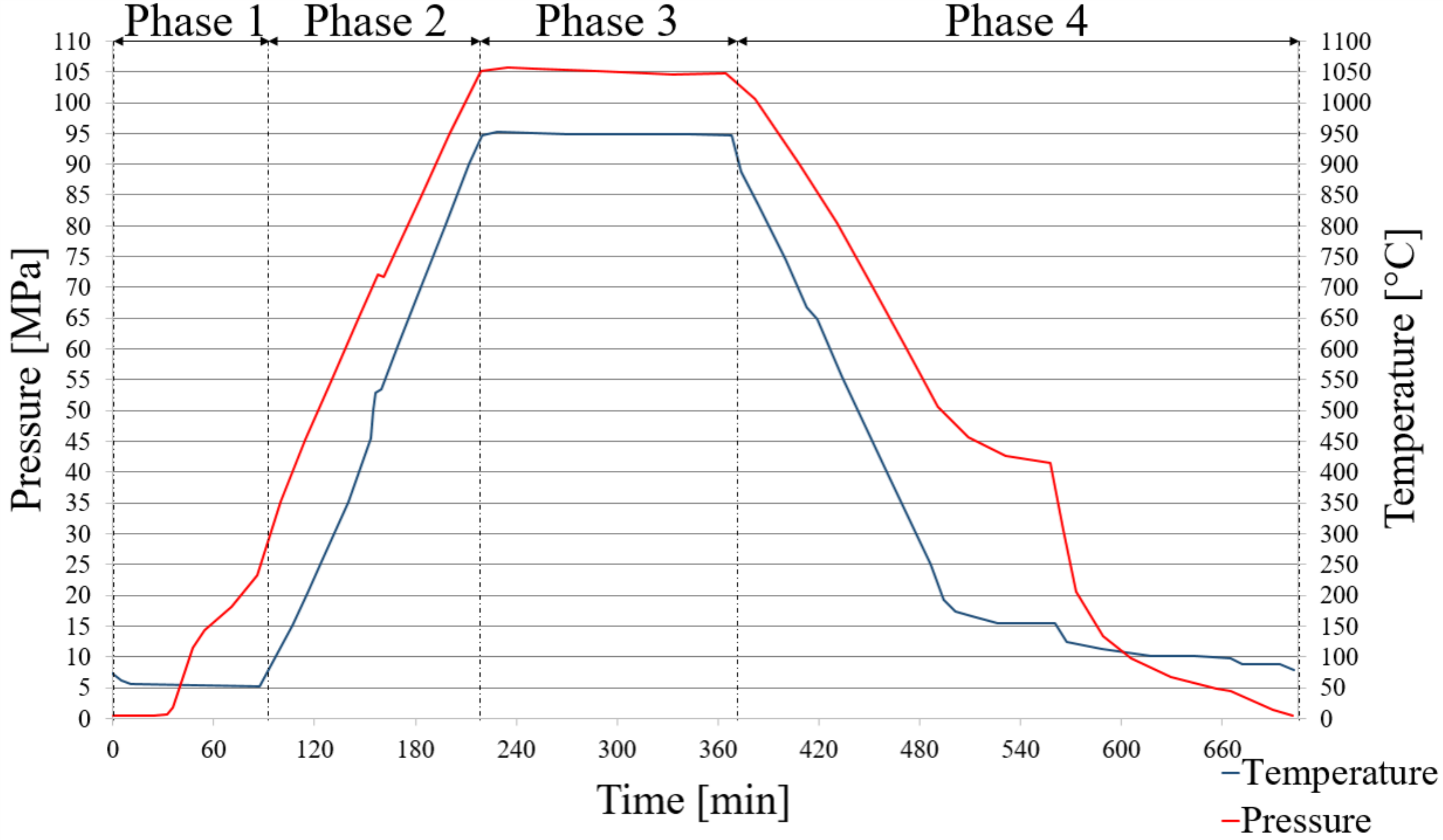}
    \caption{Temperature and pressure over time of the HIP cycle performed for the prototype of cooling plate of the SPS beam dump}
    \label{fig:HIPCycle}
\end{figure}

As shown in Figure~\ref{fig:HIPCycle}, the HIP cycle phases and the parameters chosen are the following:
\begin{itemize}
    \item Phase 1,“$Starting-phase$”: the pressure is raised up to 22.5~MPa and the temperature is kept constant to “ambient” value (around 50~°C). The parameters of this phase might change as they depend on the specific HIP unit used.
    \item Phase 2, “$Heating-phase$”: linear pressure and temperature increase up to ~105~MPa and 950~°C. The temperature increases at a rate of $\sim$ 5~°C/min.
    \item Phase 3, “$Sustaining-phase$”: the following parameters of temperature and pressure are kept over 180~min:
    \begin{itemize}
        \item Temperature: 950~°C
        \item Pressure: 105~MPa
    \end{itemize}
    \item Phase 4, “$Cooling-phase$”: pressure and temperature are decreased to ambient value. The temperature decreases by natural cooling at a rate of $\sim$ 5~°C/min.
\end{itemize}

\subsection{Ultrasonic testing examination and capsule opening}
\begin{figure}
    \centering
    \includegraphics[width=0.41\textwidth]{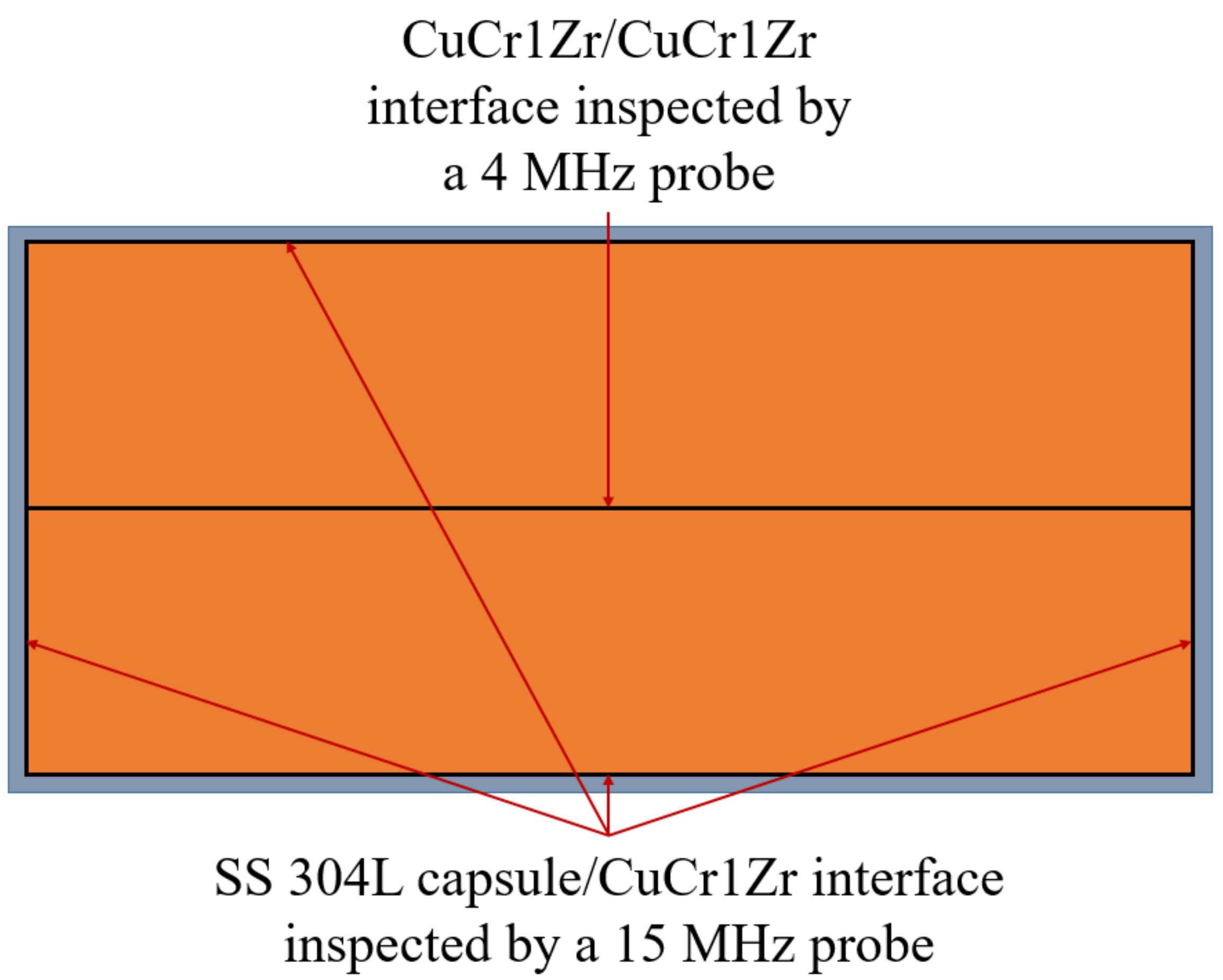}
    \caption{Interfaces inspected by ultrasounds after HIP cycle}
    \label{fig:UT}
\end{figure}

Inspection by Ultrasonic Testing (UT) is a preliminary and fast method to check if the parts have achieved an intimate contact after the HIP cycle. The interfaces that can be easily inspected are the SS~304L capsule/CuCr1Zr and CuCr1Zr/CuCr1Zr as shown in Figure~\ref{fig:UT}. If the HIPing process is successful, the different parts of the assembly are in intimate contact and there are very few or no reflections at the interfaces. Specifically, at the interface between the SS~304L capsule and CuCr1Zr, the ultrasound beam will only be reflected in a small portion due to the slight difference in acoustic impedance of the joined materials. This interface was inspected with an ALPHA2, 15 MHz and 6 mm diameter probe. With a gain of 40 dB and the speed of sound in stainless steel of ~5800 m/s, the reflection obtained was less than 5\%. The interface between CuCr1Zr/CuCr1Zr, given its bigger depth, was inspected with the lower frequency SEB4 probe of 4 MHz and 20 mm diameter. The only reflection obtained was the back-wall echo corresponding to the thickness of the assembly. This inspection was carried out by setting a gain of 60 dB and the speed of sound in copper of about 4800 m/s. The ultrasound beams generator was an USM Go+. The probes and the UT generator were provided by General Electric. After the UT inspections, the capsule was removed by machining.

\subsection{HIP diffusion bonded interface examination}
The interfaces were inspected with optical and electronic microscopes to assess the diffusion bonding quality. The equipment used for the examinations consisted in:
\begin{itemize}
    \item Digital microscope KEYENCE VHX 6000;
    \item Scanning Electron Microscope (SEM), field emission gun FEG Sigma (ZEISS) with InLens (Secondary  Electron), Everhart-Thornley Secondary Electron (SE2), and back-scattered electron (AsB) detectors for imaging;
    \item $50 mm^2$ X Max Electron dispersive X-ray spectroscopy (EDS) detector (Oxford), AZTEC \mbox{software}.
\end{itemize}

\begin{figure}
    \centering
    \includegraphics[width=0.5\textwidth, height=7cm]{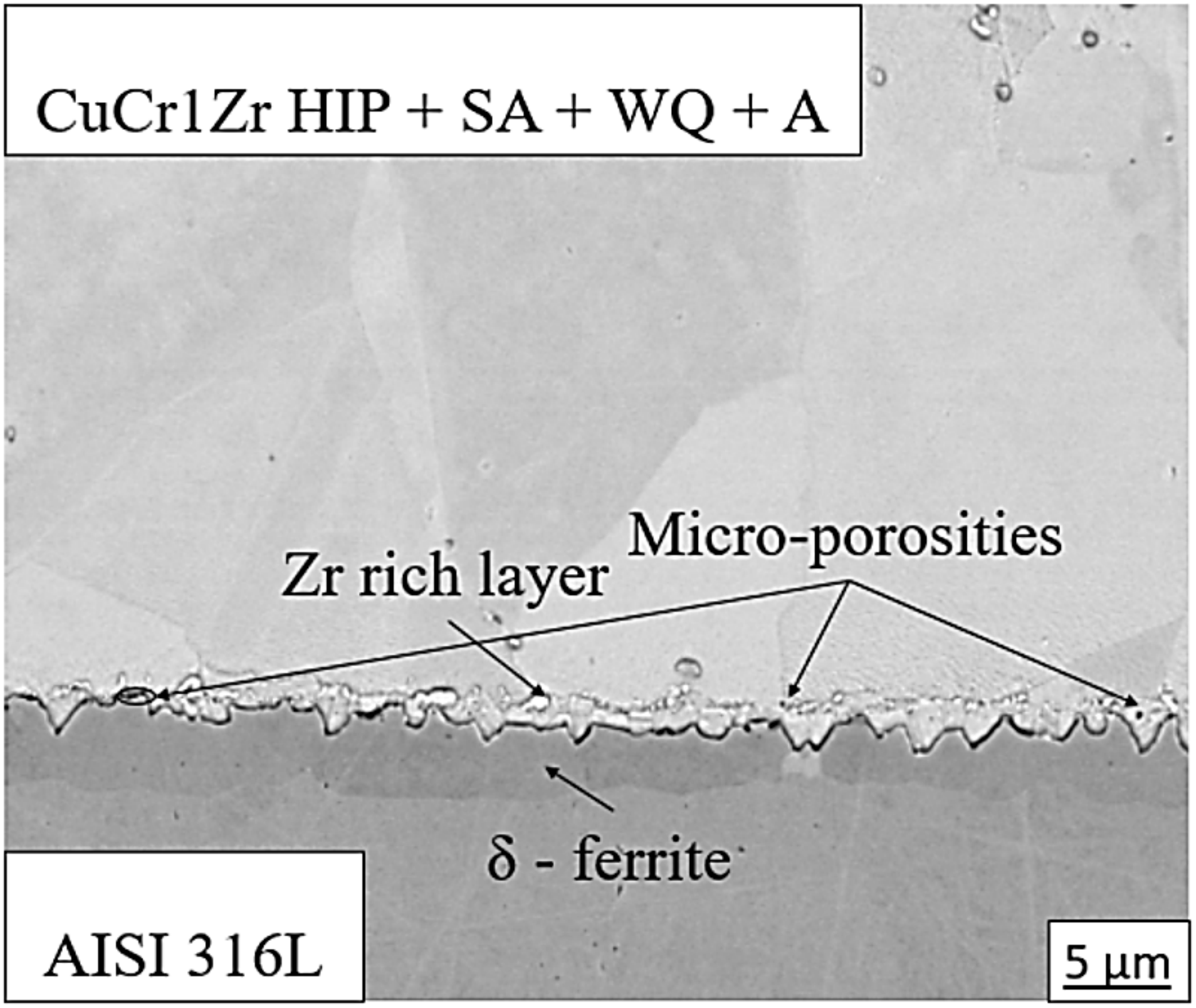}
    \caption{SEM image at 3000 X and 15kV accelerating voltage. A continuous $\delta$-ferrite layer is visible on the SS side, whereas micro porosities due to the Kirkendall effects and Zr inclusions are distinguishable on the CuCr1Zr side.}
    \label{fig:SEMInt}
\end{figure}

Figure~\ref{fig:SEMInt} shows that the interface CuCr1Zr - SS~316 L, is free from major defects. The bonding zone shows a low amount of micro-porosity on the CuCr1Zr side, most probably corresponding to the Kirkendall effect~\cite{SS_reactions} due to the different diffusion coefficients between the bonding materials, a first indicator that the diffusion took place.

\begin{figure}[htbp]
    \centering
    \includegraphics[width=0.5\textwidth, height=6cm]{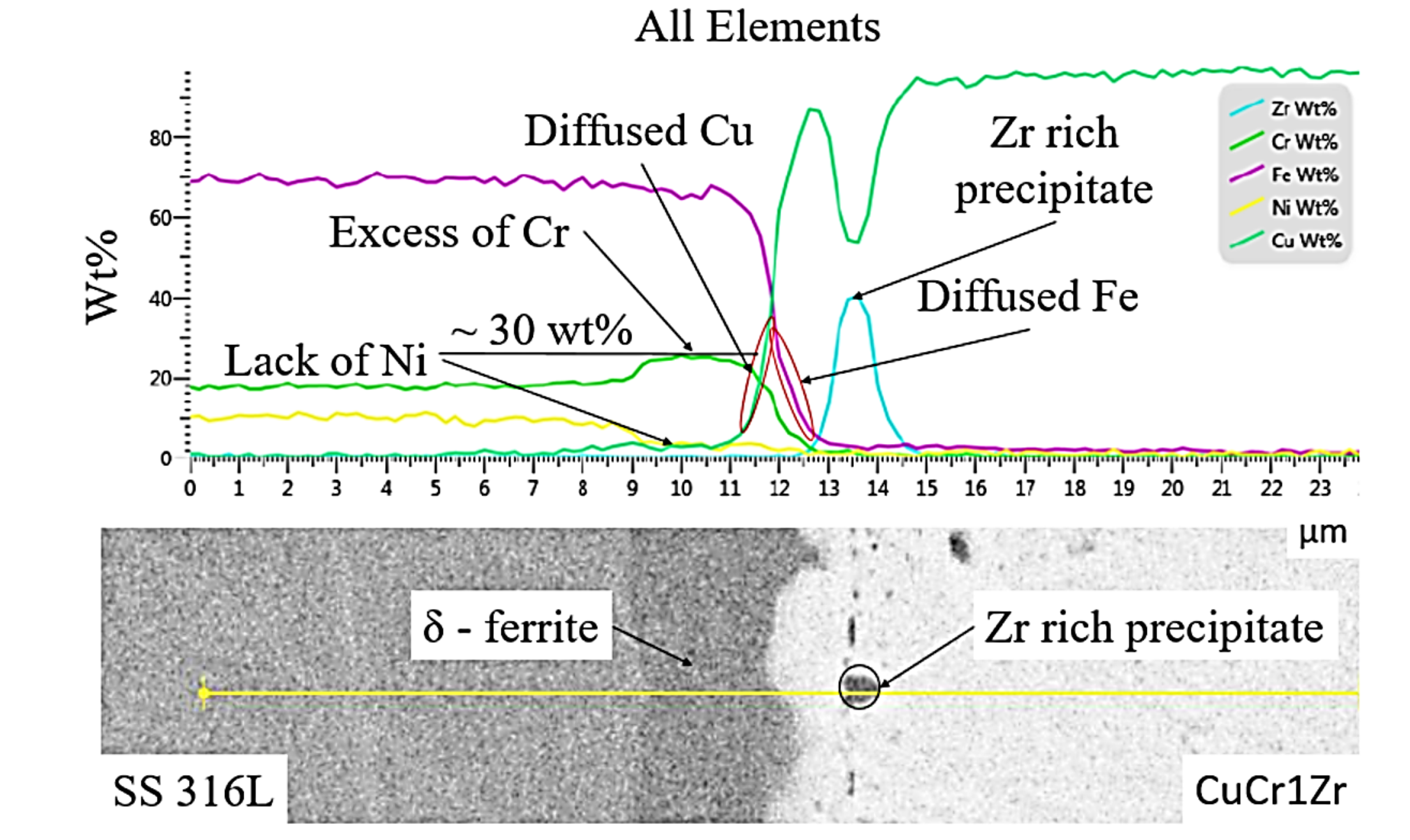}
    \caption{EDX elemental line-scans acquired at 20 keV accelerating voltage in a direction perpendicular to the interface in order to assess the diffusion phenomena. The “x” axis represents in $\mu$m the yellow line at the bottom of the picture and the “y” axis gives the concentration of the chemical elements along this line. On the SS 316L side, there is a clear excess of Cr up to ~30 Wt\% and a lack of Ni with some diffused Cu. On the CuCr1Zr side, it can be distinguished a peak corresponding to a Zr inclusion and some diffused Fe.}
    \label{fig:EDXInt}
\end{figure}

\begin{figure}
    \centering
    \includegraphics[width=0.4\textwidth]{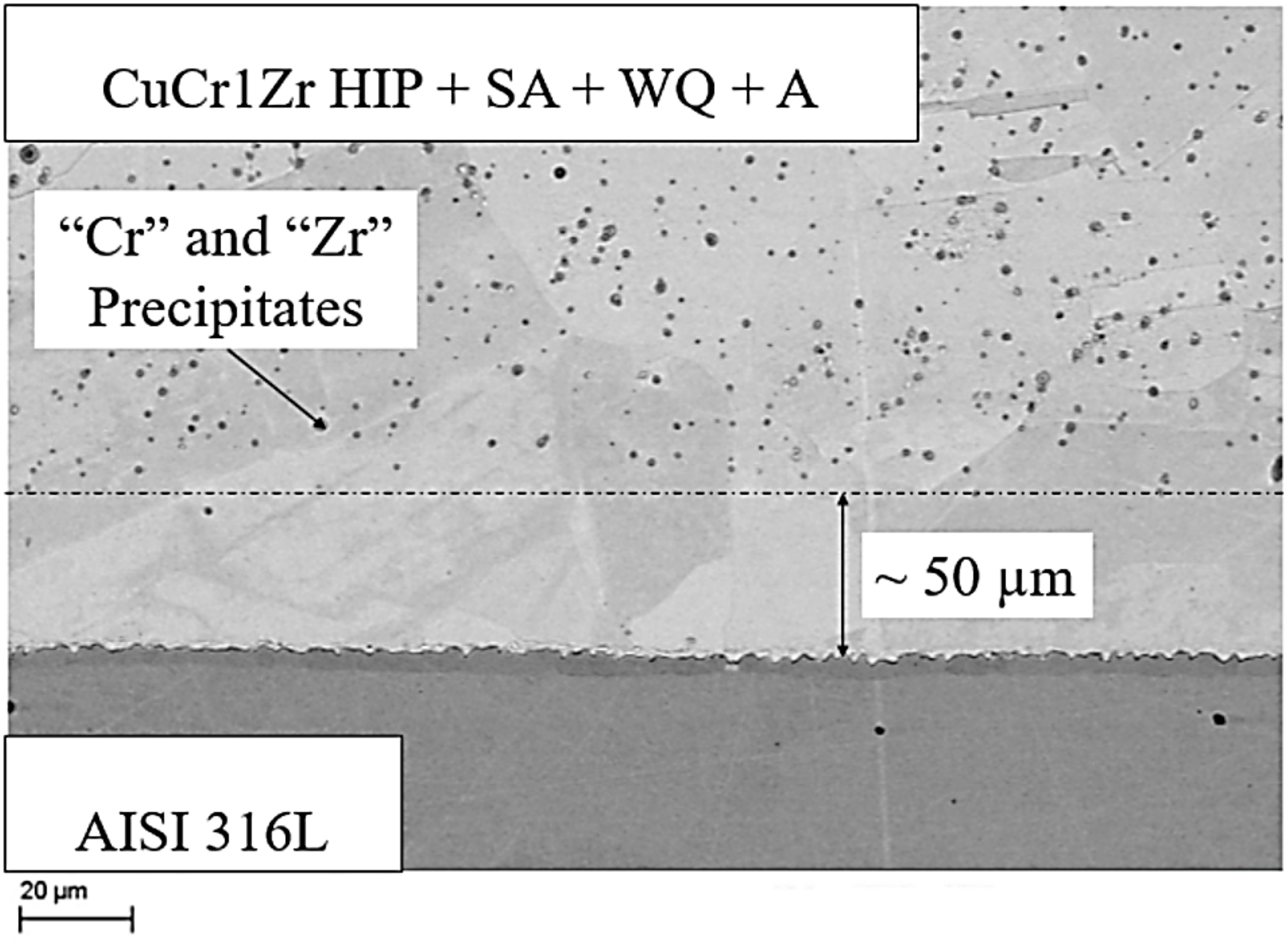}
    \caption{SEM image 500X of the CuCr1Zr-SS 316L interface. Dark “Cr” and “Zr” precipitates consolidated during aging are visible on the CuCr1Zr side}
    \label{fig:SEMInt2}
\end{figure}

As shown by Figure~\ref{fig:EDXInt}, several EDX analyses of the interface CuCr1Zr – SS~316L revealed a general trend: a few micrometres Cr rich and Ni poor layer with some diffused Cu is present on the steel side whereas Zr rich precipitates with some diffused Fe are on the CuCr1Zr side~\cite{PS_Prot}. The darker continuous layer of about 5~µm thickness visible on the steel side is $\delta$-ferrite. According to the Schaeffler diagram for stainless steel, nickel loss at the SS~316L/CuCr1Zr interface, Cr diffusion to SS~316L, plus Cr enrichment up to ~ 30 wt\% are responsible for the formation of a fully ferritic layer at the joint. Figure~\ref{fig:SEMInt2} also shows large dark Cr and Zr precipitates consolidated during the aging process which diminish in a layer of about 50~µm close to the SS316L /CuCr1Zr interface.

\begin{figure}
    \centering
    \includegraphics[scale=0.4]{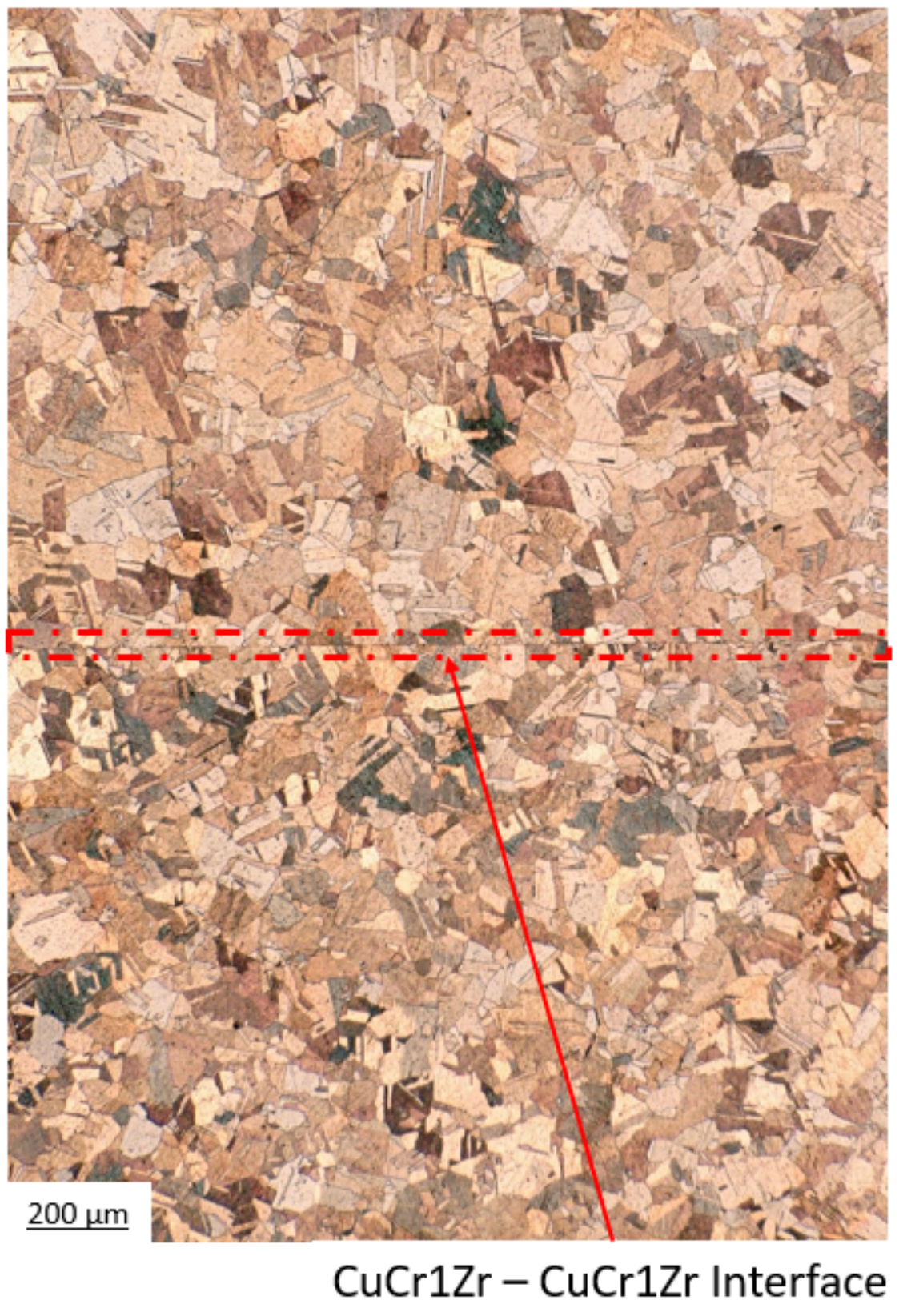}
    \caption{Optical microscopy of a sample containing the HIP diffusion bonded CuCr1Zr-CuCr1Zr interface. The surface inspected has been mechanically polished and chemically etched.}
    \label{fig:OptInt}
\end{figure}

Figure \ref{fig:OptInt} shows an optical microscopy of the interface CuCr1Zr-CuCr1Zr, which appears clearly discernible, homogeneous and without significant voids population. Interfacial grain boundaries lie in the plane of the joint.

\subsection{Thermal conductivity measurements}
\begin{figure}[h!]
    \centering
    \includegraphics[width=0.5\textwidth]{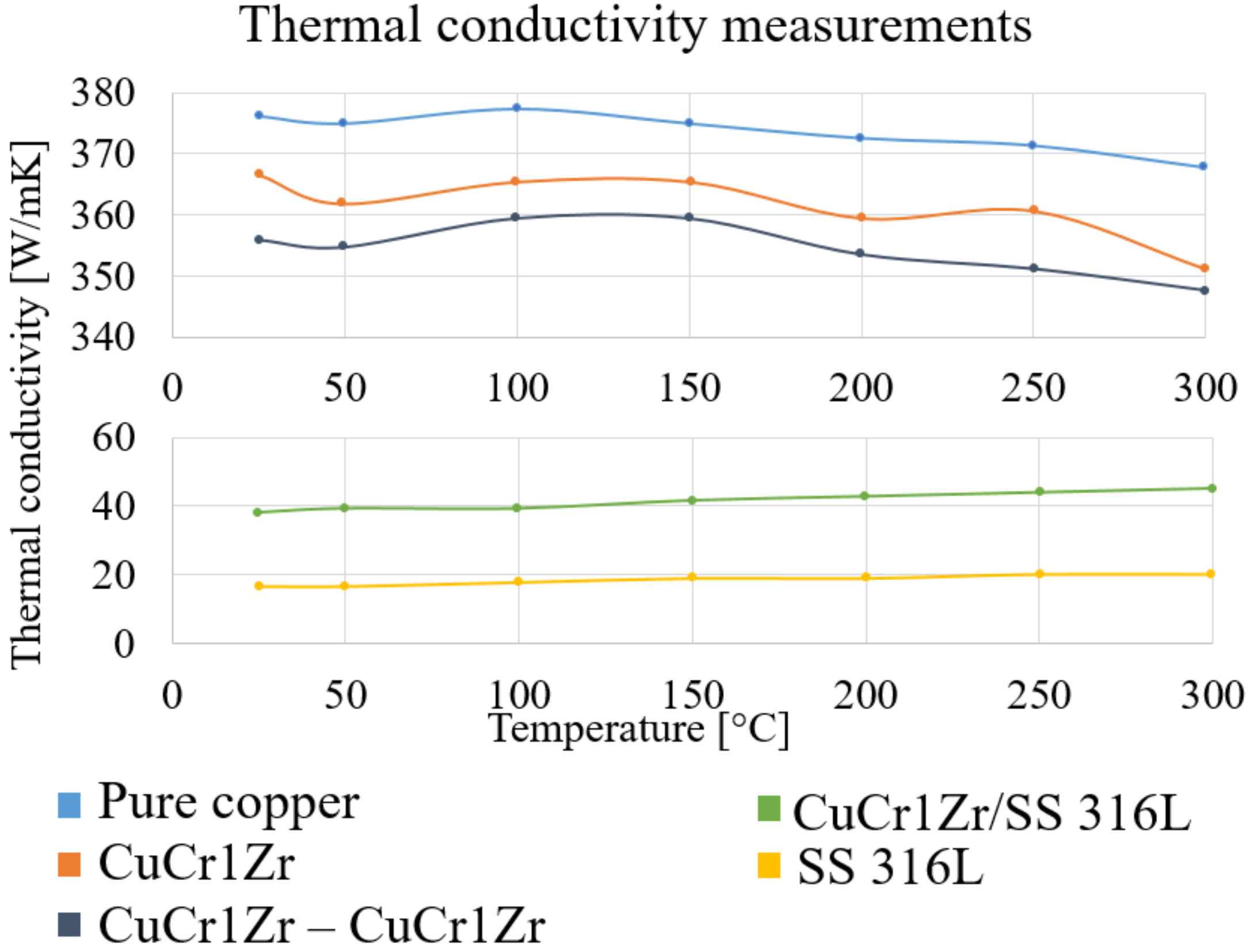}
    \caption{Thermal conductivity measurements at the materials’ bulks and interfaces.}
    \label{fig:TC}
\end{figure}

Figure~\ref{fig:TC} shows the measurements of thermal conductivity, from room temperature to 300~°C, of pure copper, precipitation hardened CuCr1Zr, SS~316L and on samples containing their respective HIP diffusion-bonded interfaces, performed by using a NETZSCH nanoflash LFA447.

At the interface CuCr1Zr-CuCr1Zr (dark blue line), the thermal conductivity is as high as that measured in the bulks of precipitation hardened CuCr1Zr (orange line) and pure copper (light blue line), with values ranging between $350 \pm 10 Wm^{-1}K^{-1}$ and $360 \pm 10 Wm^{-1} K^{-1}$. Similarly, the thermal conductivity at the interface CuCr1Zr-SS~316L (green line) and the bulk of SS~316L (yellow line) are close and fall in the range between $38 \pm 1 Wm^{-1}K^{-1}$ and $45.0 \pm 1.4 Wm^{-1}K^{-1}$. These results indicated that at the HIP diffusion bonded interfaces the values are limited by the material with the lowest thermal conductivity in the joint, however higher than the latter (see e.g. CuCr1Zr-SS~316L interface).

\subsection{Mechanical strength measurements}
\begin{figure}[h!]
    \centering
    \includegraphics[width=0.3\textwidth]{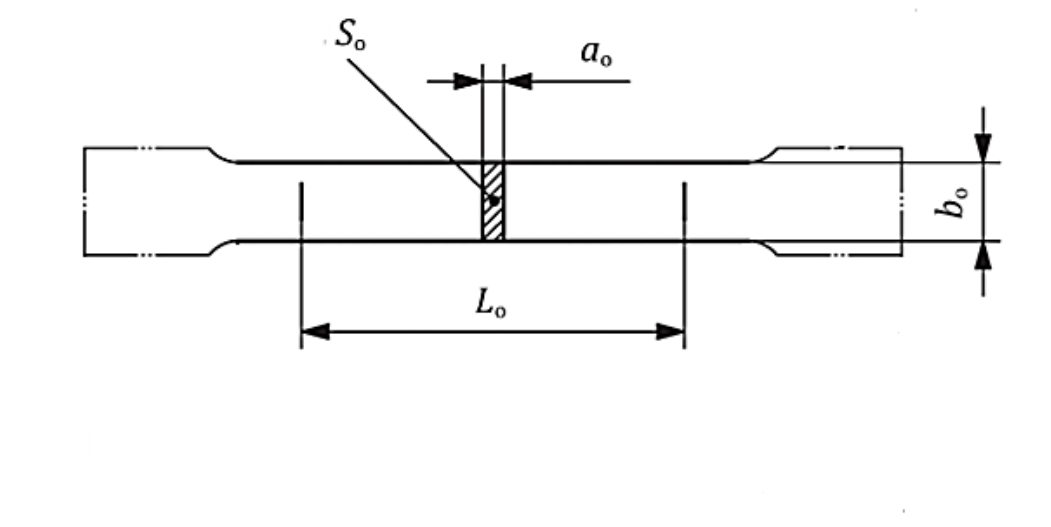}
    \caption{Tensile samples dimensions}
    \label{fig:TSample}
\end{figure}

Tensile tests at room temperature, according to the DIN EN ISO 6892-1, of the HIP diffusion bonded interfaces were performed in order to quantify their mechanical strengths. Samples made of CuCr1Zr and SS~316L as well as containing the HIP diffusion bonded interfaces were prepared by electro-erosion. The initial cross section dimensions and gauge length are the following (Figure~\ref{fig:TSample}):

\begin{itemize}
    \item Width: $b_0 = 1.5 mm$
    \item Thickness $a_0 = 2 mm$
    \item Cross-sectional area: $S_0 = 3 mm^{2}$
    \item Gauge length: $L_0 = 4.6 mm$
\end{itemize}

\begin{figure}[h!]
    \centering
    \includegraphics[width=0.5\textwidth]{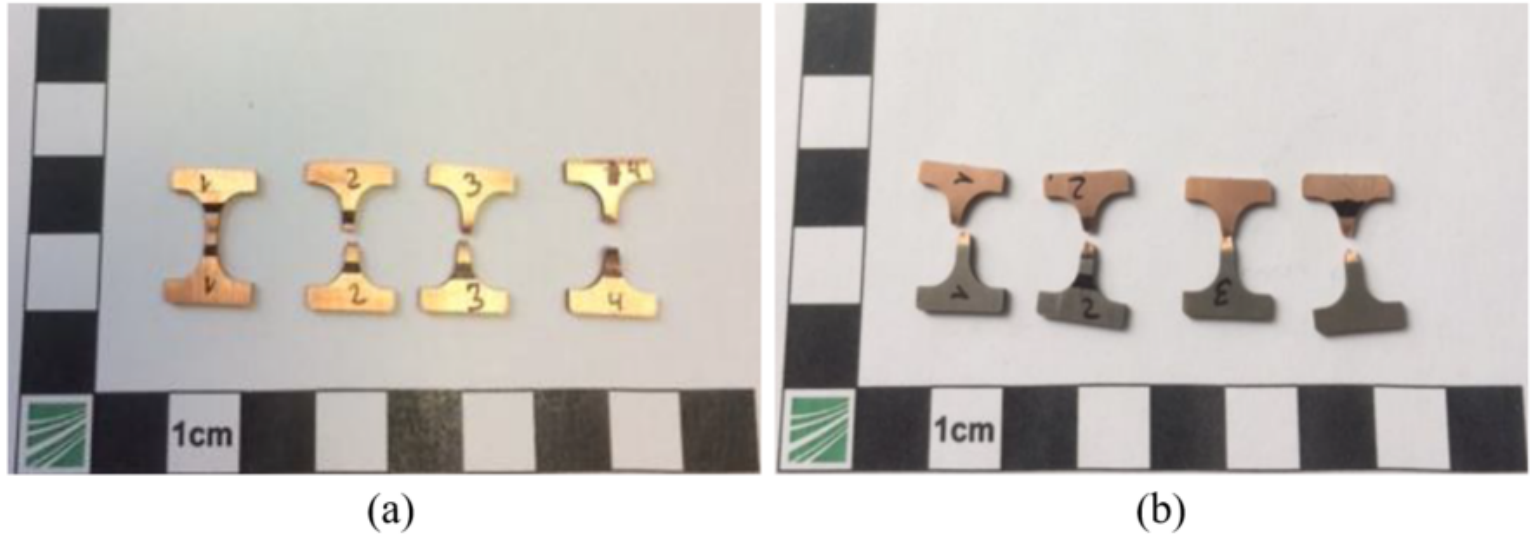}
    \caption{Tensile samples containing (a) the CuCr1Zr-CuCr1Zr interface and (b) the CuCr1Zr-SS~316L interface.}
    \label{fig:TSampleInt}
\end{figure}

Figure~\ref{fig:TSampleInt} shows the tested tensile samples containing the CuCr1Zr-CuCr1Zr and CuCr1Zr-SS~316L interfaces within the gauge length.

\begin{figure}
    \centering
    \includegraphics[width=0.45\textwidth]{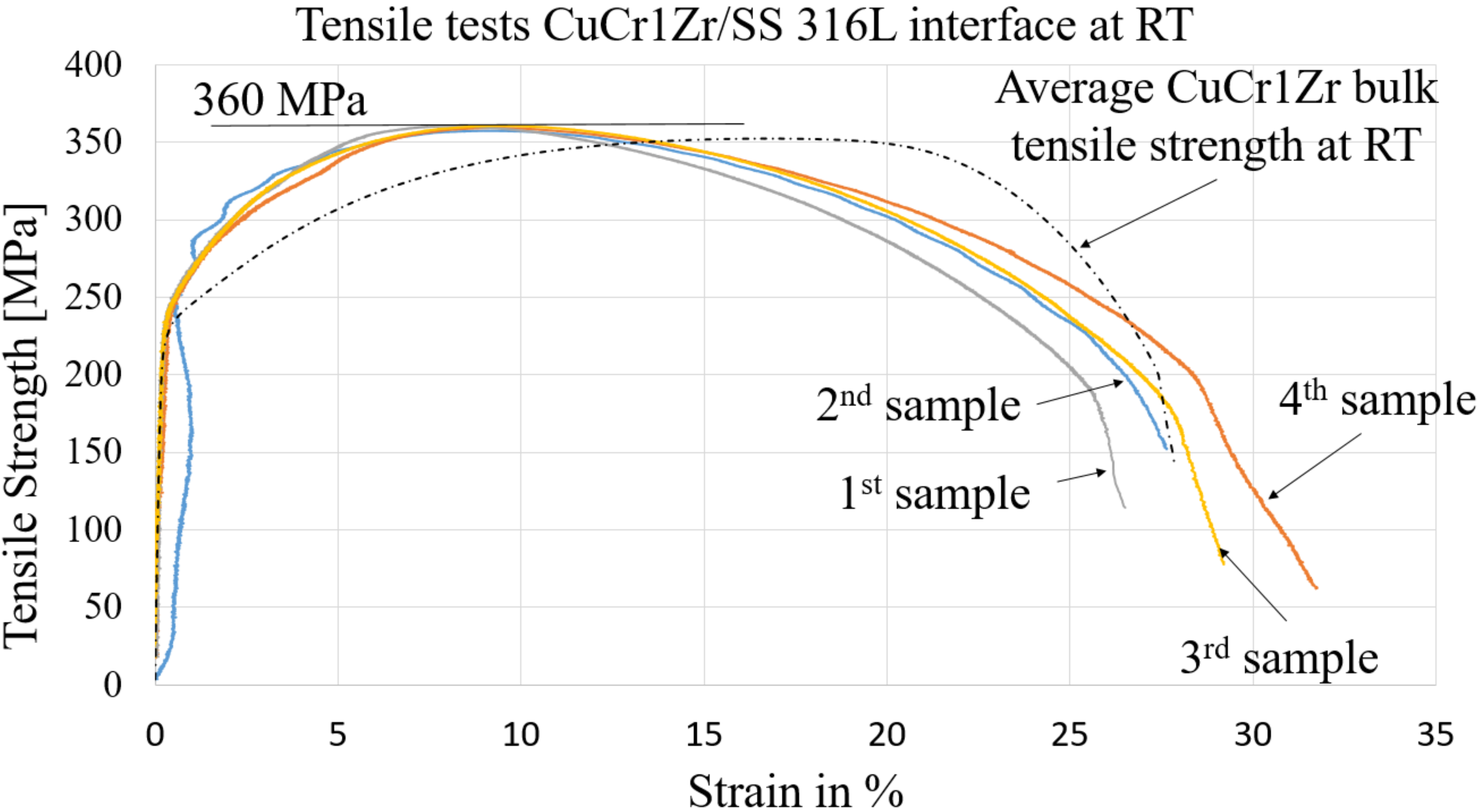}
    \caption{Room temperature tensile stress-strain curves for CuCr1Zr-SS~316L specimens and bulk hardened CuC1Zr as a reference.}
    \label{fig:SigmaStrainCuSS}
\end{figure}

\begin{figure}[H]
    \centering
    \includegraphics[width=0.45\textwidth]{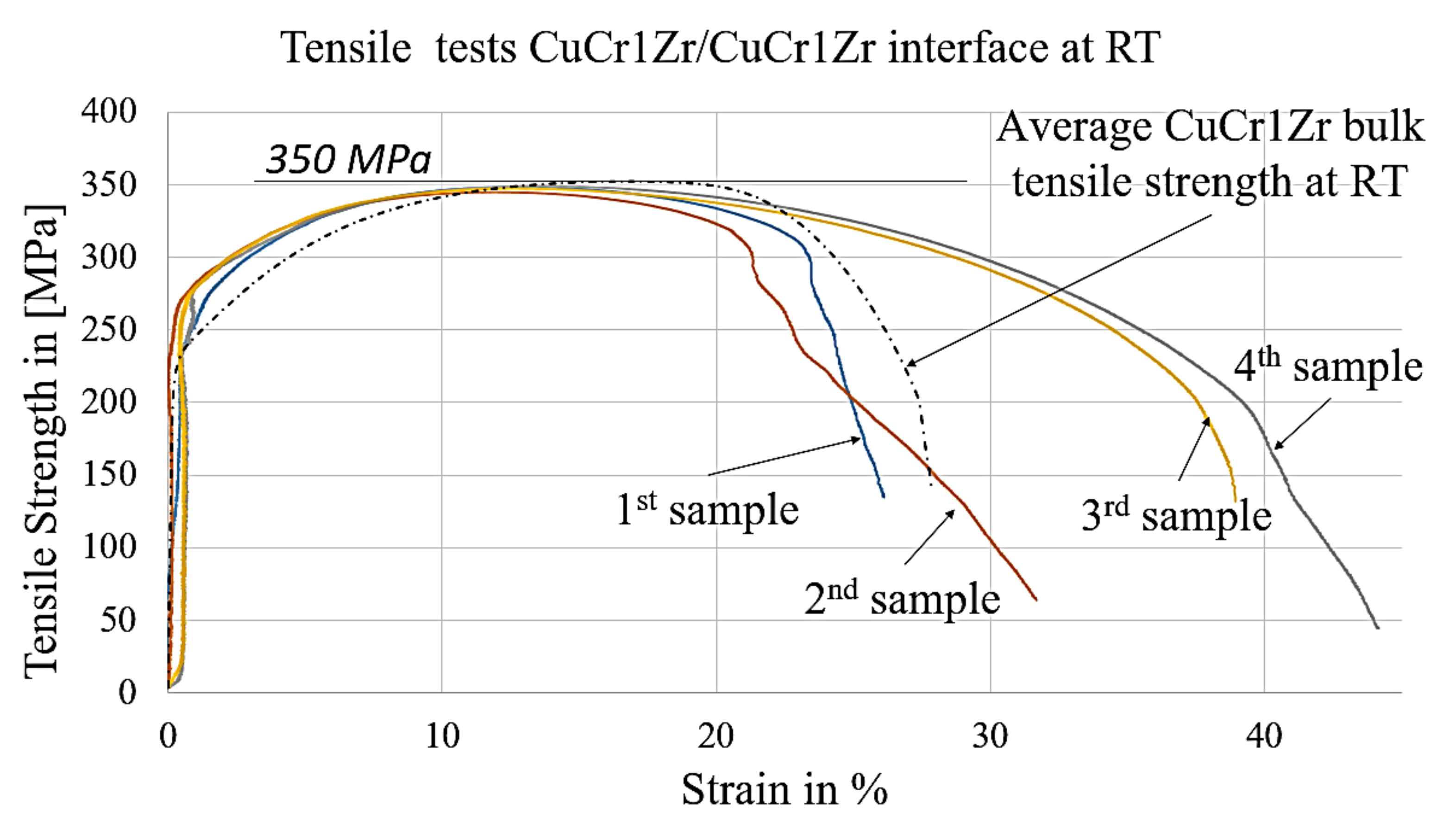}
    \caption{Room temperature tensile stress-strain curves for CuCr1Zr-CuCr1Zr specimens and bulk hardened CuCr1Zr.}
    \label{fig:SigmaStrainCuCu}
\end{figure}

For the CuCr1Zr-SS~316L samples, the fracture appears to be ductile and it is not at the interface but within CuCr1Zr. The measured mean tensile strength of 360 $\pm$ 2 MPa is comparable with precipitation hardened CuCr1Zr, as it is shown by the dotted line in Figure~\ref{fig:SigmaStrainCuSS}. Similarly, Figure~\ref{fig:SigmaStrainCuCu} shows the results of the tensile samples containing the CuCr1Zr-CuCr1Zr interface. The mean measured tensile strength of 348 $\pm$ 2 MPa is very close to the bulk precipitation hardened CuCr1Zr of 352 $\pm$ 3 MPa. The presence of a diffusion bonded interface does not seem to affect the mechanical strength, which is, instead, limited by the lowest one of the joining materials. Nevertheless, the samples containing the CuCr1Zr/SS~316L interface exhibit a lower elongation at necking with respect to the bulky CuCr1Zr ones. This difference is smaller for the CuCr1Zr/CuCr1Zr samples. However, the total elongations are comparable and in most cases close to ~30\%. Table~\ref{Table5} summarizes the mean values of tensile strengths.

\begin{table}[h!tbp]
    \centering
    \caption{Mean tensile strengths of the specimens with the interfaces CuCr1Zr/CuCr1Zr, CuCr1Zr/SS~316L and for specimens of CuCr1Zr without interface.}
    \begin{tabular}{|c|c|}
    \hline
         Samples & Tensile strength [MPa]  \\
         \hline
         CuCr1Zr/SS 316L & $360 \pm 2$ \\
         \hline
         CuCr1Zr/CuCr1Zr & $348 \pm 2$ \\
         \hline
         Bulk CuCr1Zr & $352 \pm 3$\\
         \hline
    \end{tabular}
    \label{Table5}
\end{table}

\subsection{Cooling performance test bench}
\begin{figure}[h!]
    \centering
    \includegraphics[width=0.5\textwidth]{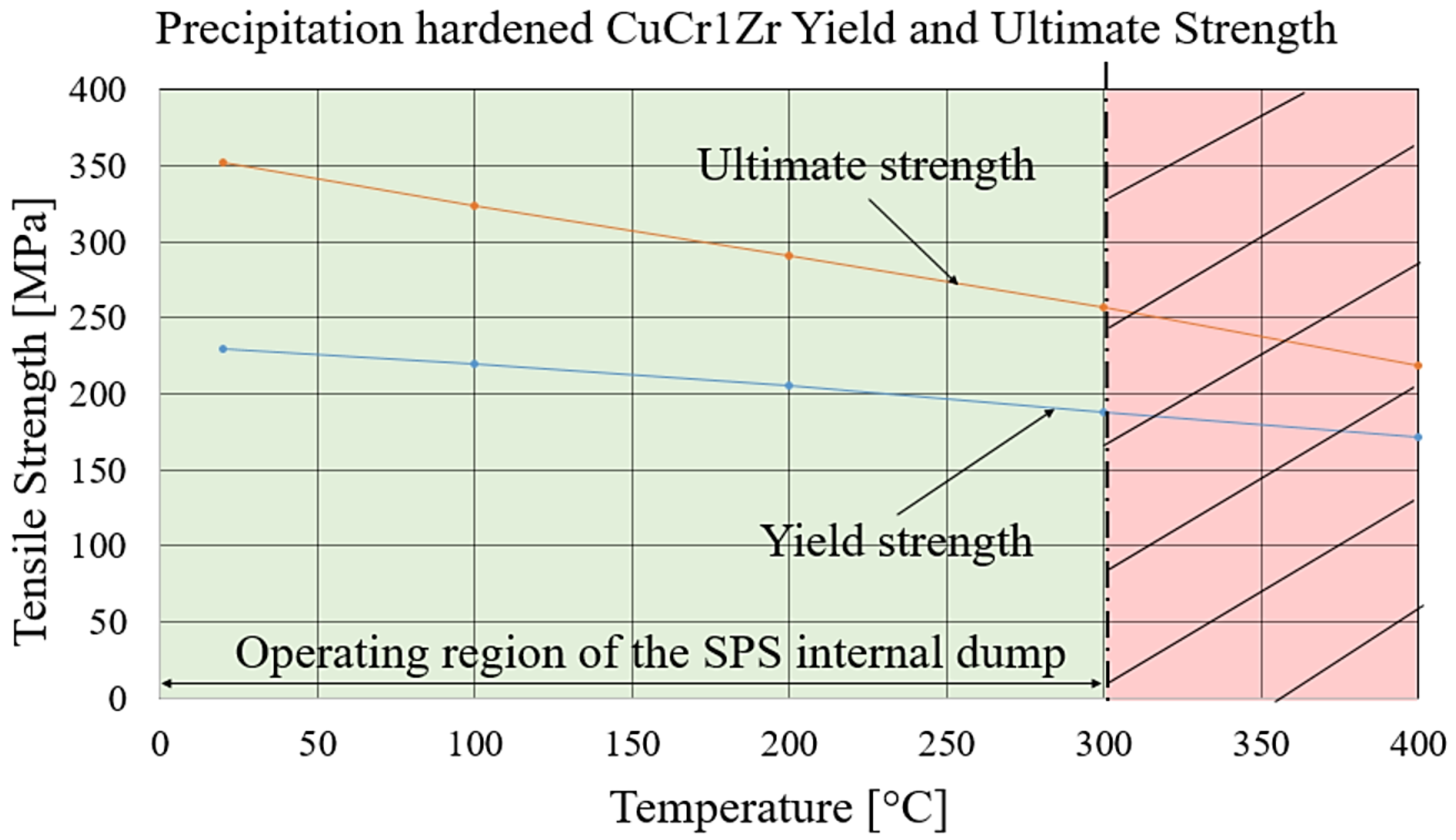}
    \caption{Yield and ultimate strength of precipitation hardened CuCr1Zr from room temperature to 400°C. Every CuCr1Zr component of the SPS internal dump shall operate at a temperature always below the limit of 300°C.}
    \label{fig:SigmaTempCu}
\end{figure}

In order to assess the ability of the HIP manufactured prototype to evacuate the thermal power deposited in the most critical regions of the SPS internal dump’s cooling plates and in the worst operational scenario, as well as to verify that the cooling performance all over the prototype was homogeneous, a dedicated test bench was designed.
In order to validate the first point, the cooling plate prototype needed to be able to reach a steady state with an average temperature well below 300 °C. Indeed, as Figure~\ref{fig:SigmaTempCu} shows, starting from 300~°C the measured yield strength of precipitation hardened CuCr1Zr becomes too low for a robust and reliable operation of the SPS internal dump.

As for the homogeneity of the cooling performance, the amount of power evacuated by the two tubes as well as the cooling rates measured at different locations of the prototype plate shall be the same within the measurement errors.

\subsubsection{Experimental setup}
\begin{figure}
    \centering
    \includegraphics[width=0.5\textwidth]{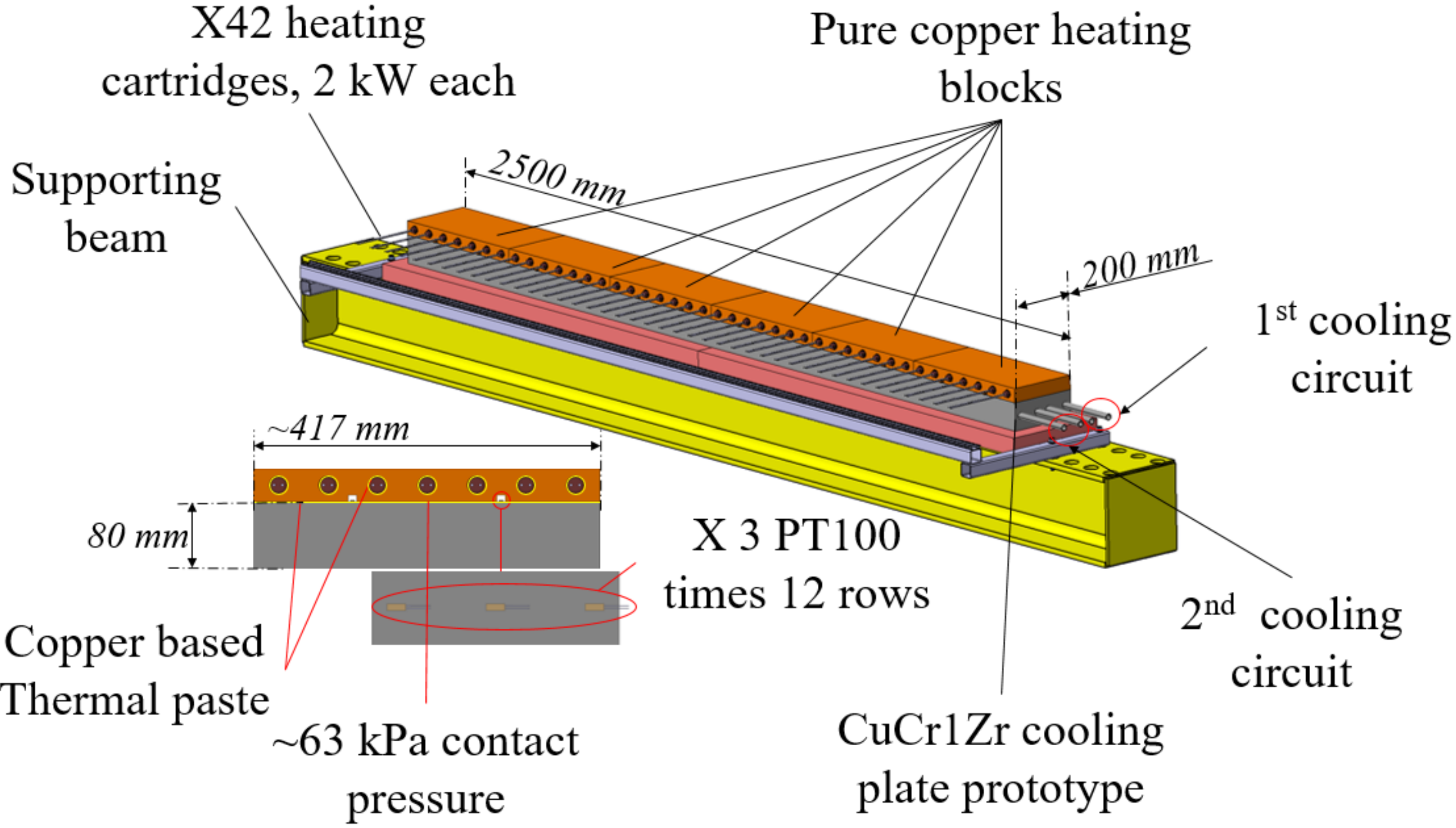}
    \caption{Components of the cooling test bench and the respective configuration.}
    \label{fig:TestBench}
\end{figure}

The test bench is made of four main systems/part: the heating and cooling systems, the monitoring system and the part tested. The heating system reproduces the thermal power deposited by the particle beam. The cooling system extracts this power in the same way the actual system would operate on the dump's cooling plates in operation and finally the monitoring system gives feedback about the temperatures and the inlet/outlet flow rates of the part being tested.
Figure~\ref{fig:TestBench} shows the main test bench components used to assess the cooling performance of the HIP manufactured cooling plate prototype. 
Specifically, the thermal power deposited by the particle beam is reproduced by six oxygen-free high thermal conductivity (OFHC) copper heating blocks. Every heating block has 7 electrical heating cartridges 2~kW each, for a total power of 84~kW.

In order to improve the heat flow, a copper-based high temperature lubricant with very good thermal conductivity (ORAPI TM900S) couples the heating cartridges with the heating blocks and these latter with the cooling plate prototype. The heating blocks are pressed against the prototype plate establishing an average calculated contact pressure of about 63~kPa. The temperature distribution at the interface heating blocks - cooling plate is monitored by 36 PT100s sensors. The whole assembly is enclosed within insulating fiberglass plates 50 mm thick.

\begin{figure}
    \centering
    \includegraphics[width=0.5\textwidth]{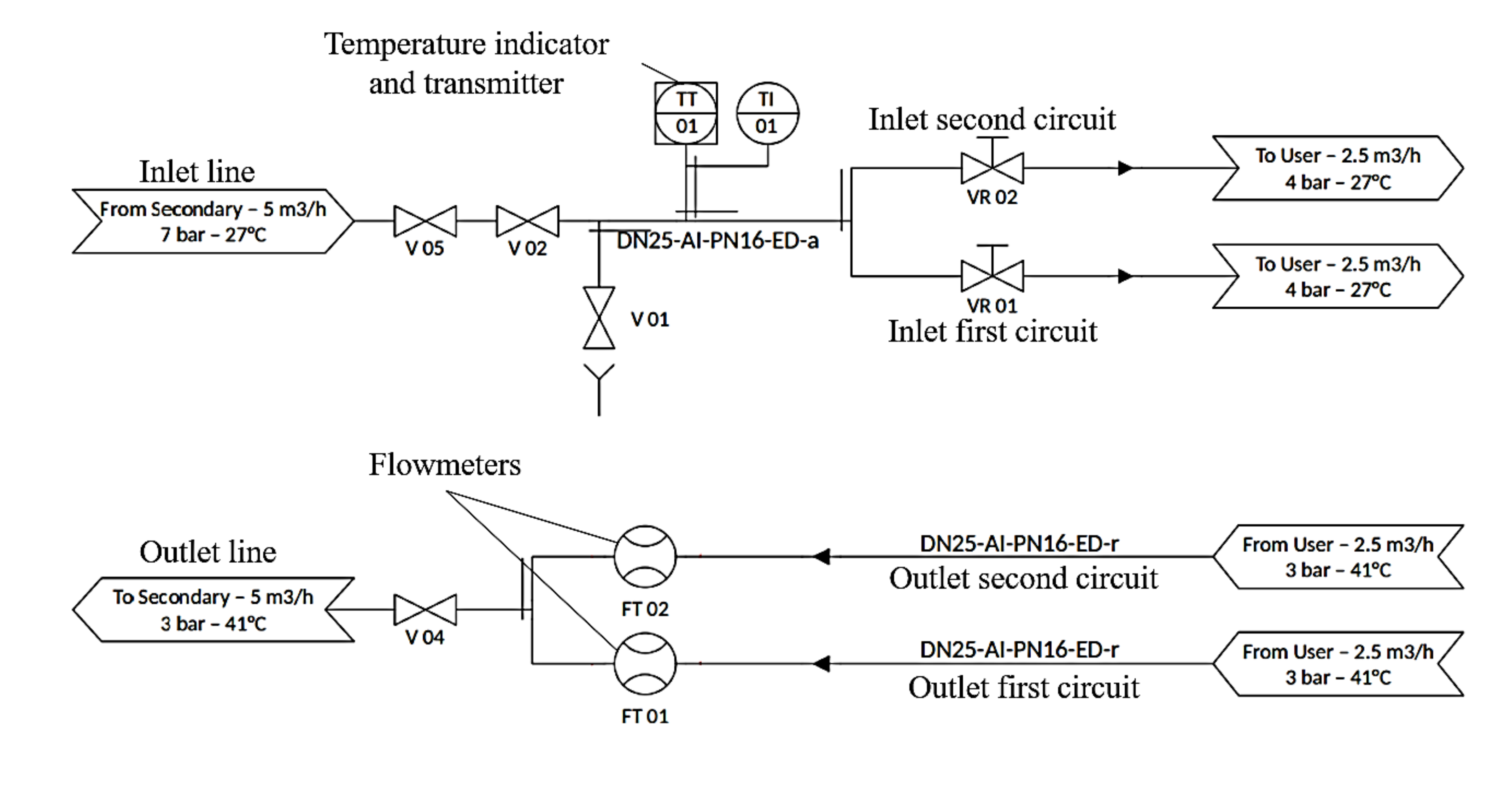}
    \caption{P\&ID cooling performance test bench}
    \label{fig:PID}
\end{figure}

The prototype plate is equipped with two cooling circuits shown in Figure~\ref{fig:TestBench}, following the same design of the SPS internal dump cooling loop. Figure~\ref{fig:PID} shows the P\&ID of the circuits. An inlet flow rate of  5 $m^{3}/h$ and  7 $bar$  of demineralized water is equally split in the two circuits of the prototype. At the inlet, the temperature is kept constant at 27~°C and it is monitored by a temperature indicator and transmitter installed directly in the line. Flowrates and temperatures at the outlet circuits are measured by two flowmeters type FC 100 LQ from Flow Vision. The electrical heating cartridges, the PT100 temperature sensors, the flowmeters and the inlet temperature transmitter are all connected to a PLC and controlled via dedicated software.

\subsubsection{Test procedure}
Every test was carried out by the following steps:
\begin{itemize}
    \item The cooling circuits are activated;
    \item The thermal power is gradually increased until reaching the testing level;
    \item The test bench is left to reach the steady state where the average cooling plate temperature and the cooling circuit' s inlet and outlets temperatures are recorded;
    \item The power is shut down and the cooling rates at different points of the plates are measured.
\end{itemize}

\subsubsection{Results}
\begin{figure}
    \centering
    \includegraphics[width=0.5\textwidth]{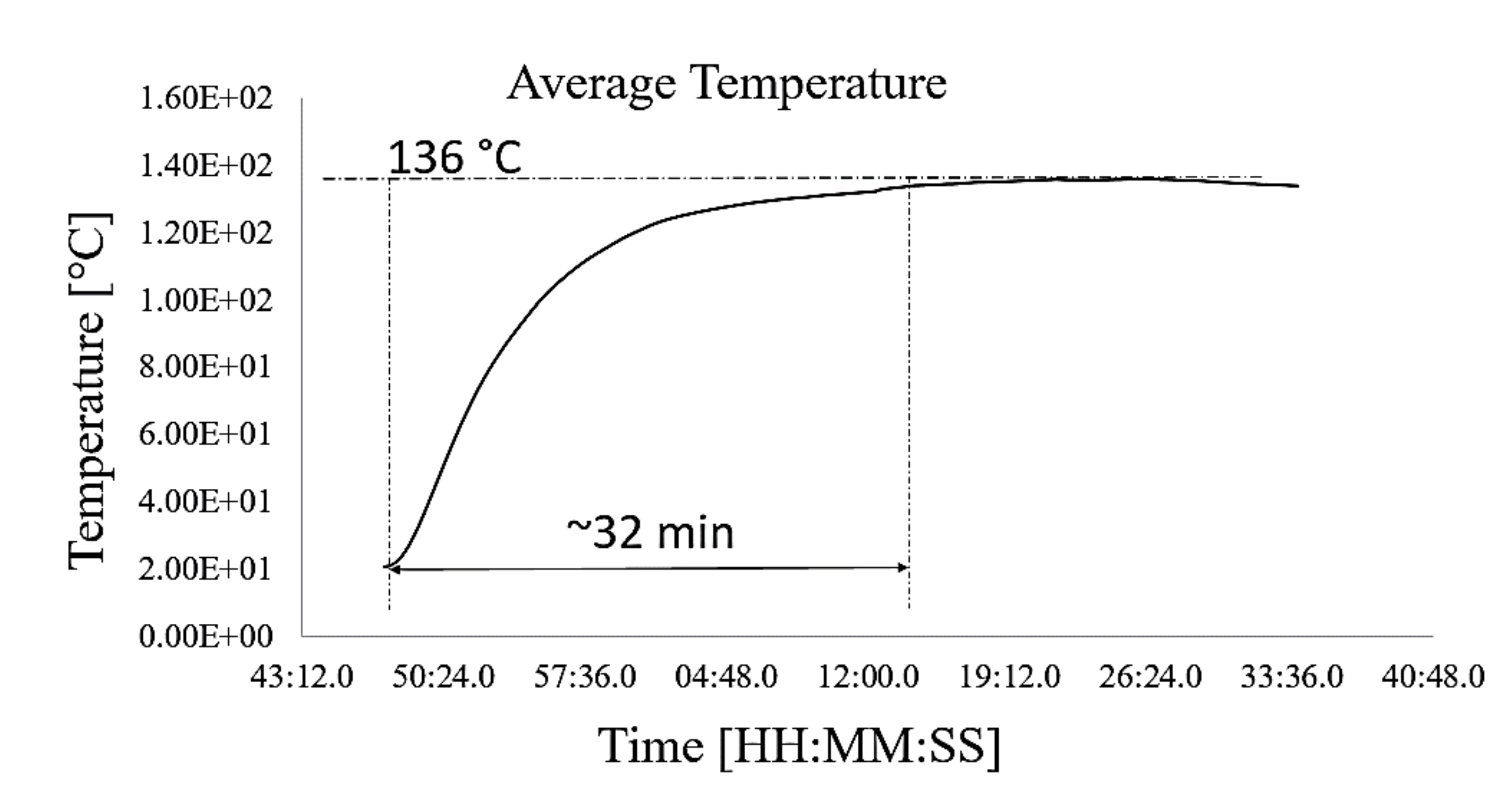}
    \caption{Average PT100s temperature over time to steady state of the heating phase.}
    \label{fig:TempTimeBench}
\end{figure}

\begin{figure}
    \centering
    \includegraphics[width=0.5\textwidth]{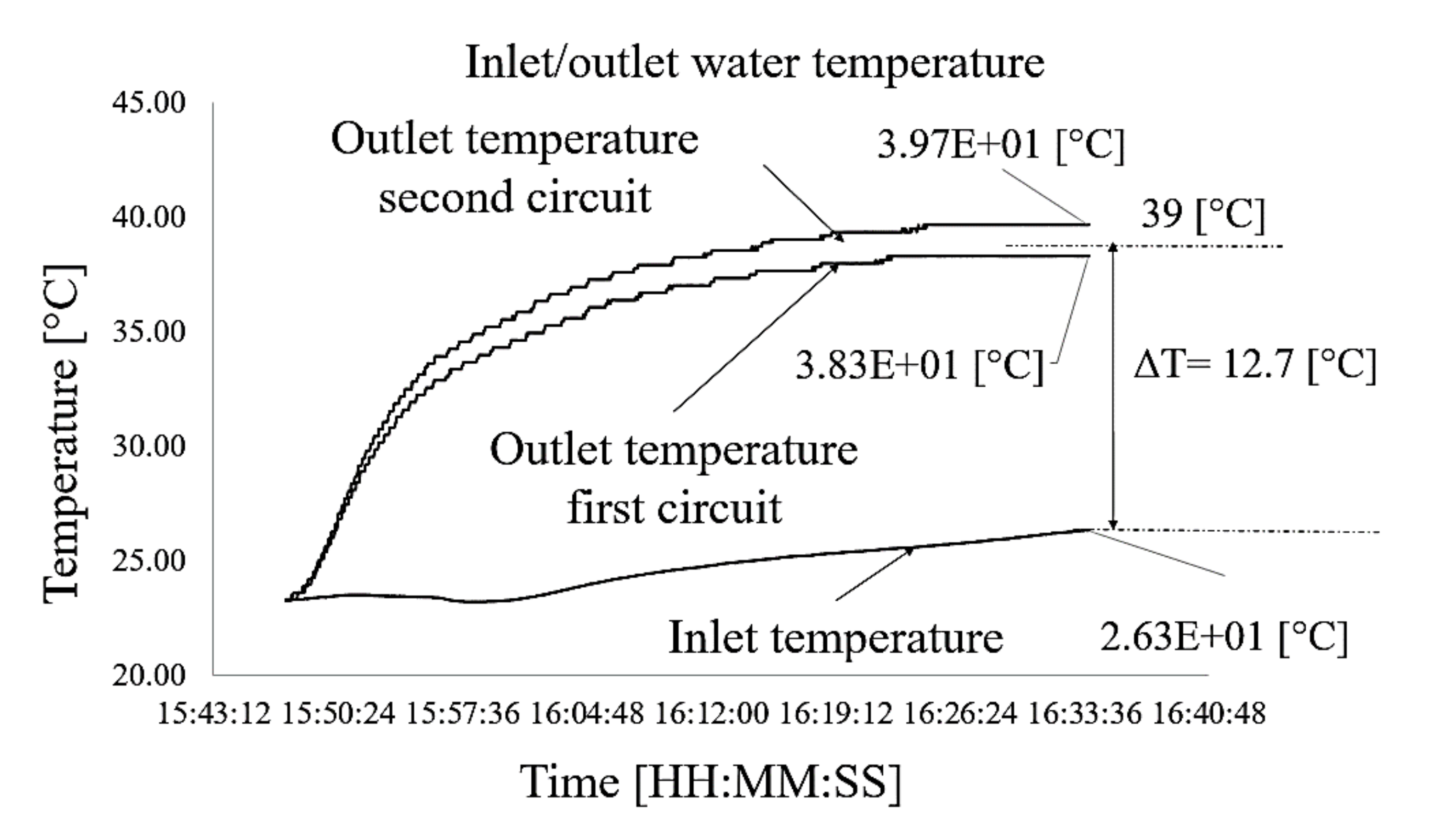}
    \caption{Inlet and outlets water temperature versus time during heating phase at $84 kW$.}
    \label{fig:TempTimeWater}
\end{figure}

\begin{figure}[h!]
    \centering
    \includegraphics[width=0.5\textwidth]{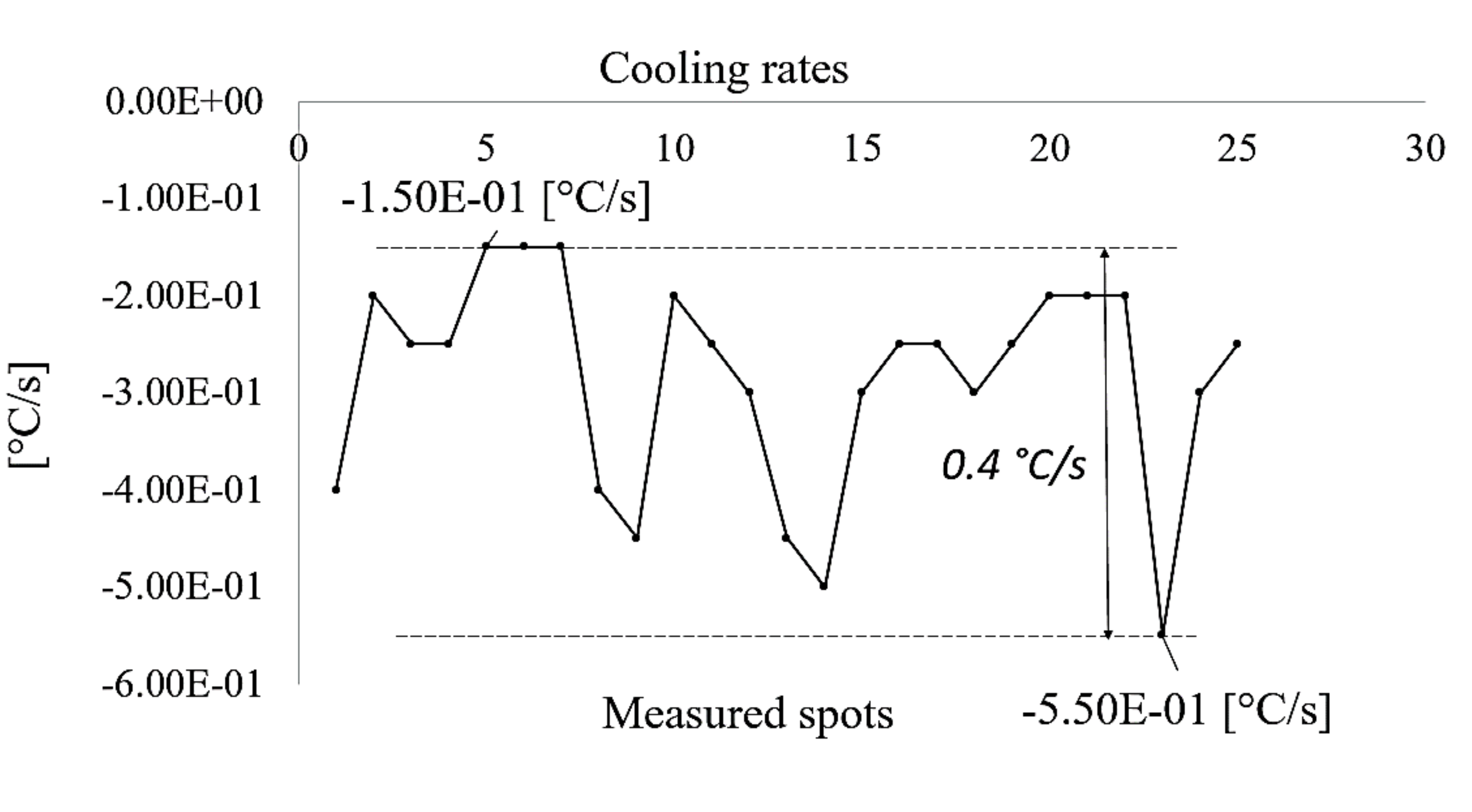}
    \caption{Maximum cooling rates measured in some points of the prototype’s surface.}
    \label{fig:CoolingRate}
\end{figure}

The highest tested thermal power was 84~kW, which is roughly equivalent to the maximum expected average power deposited in the most critical of the four plates composing the TIDVG\#5.
Figure~\ref{fig:TempTimeBench} shows the increase of average temperature of all the PT100s sensors over time, from the beginning up to the steady state. Given the cooling system in place and the amount of power tested, it took about 32 minutes to reach the steady state at an average temperature of 136~°C.
Figure~\ref{fig:TempTimeWater} shows the temperature over time at the inlet and outlets of the cooling circuits. At steady state condition, this information along with the flow rates allow to calculate the thermal power extracted by the circuits. The outlets’ temperatures difference less than 2~°C suggests that the two circuits behave in the same way and evacuate a similar amount of thermal power, precisely, 39~kW and 35~kW summing up to 74~kW. Around 10~kW are lost due to the not perfect insulation of the test bench.
When the power is shut down and the cooling plate temperature starts decreasing, the cooling rates in several points of the prototype were measured. Figure~\ref{fig:CoolingRate} shows that the maximum difference between the cooling rates is below 0.5~°C/s. This result, along with the similar thermal power evacuated, suggests that the cooling performance of the prototype are homogeneous.

\section{Discussion}
The objective of the present study was to carry out a complete characterization of the CuCr1Zr-CuCr1Zr and CuCr1Zr-SS~316L HIP diffusion bonded interfaces and to assess the cooling performance of the cooling plate prototype under the worst operational scenario of the SPS beam internal dump. In its most narrow definition, diffusion bonding is a process that produces solid-state coalescence between two clean surfaces. The process is composed of a three-stage metallurgical phases~cite{ASM}:
\begin{enumerate}
    \item The contact area grows to a large fraction of the joint area by localized deformation of the contacting surface asperities. At the completion of this stage, the interface boundary is no longer a planar interface, but consists of voids separated by areas of intimate contact. In these areas of contact, the joint becomes equivalent to a grain boundary between the grains on each surface.
    \item All of the voids in the joints shrink and most are eliminated. The interfacial grain boundary migrates out of the plane of the joint to a lower-energy equilibrium. Bonding could be considered essentially complete after this stage. 
    \item Diffusional processes caused the shrinkage and elimination of voids, now the only possible path is through the volume of the grains themselves.
\end{enumerate}
For the CuCr1Zr-SS~316L interface, there was a good contact of the surfaces and two diffusional phenomena also observed in literature~\cite{HIPing_CuSS}:

\begin{itemize}
    \item Low amount of micro-porosity on the CuCr1Zr side, most probably due to the Kirkendall effect;
    \item The Cr rich and Ni poor diffusion layer with some diffused Cu on the SS~316L side and Cr and Zr rich inclusions with some diffused Fe on the CuCr1Zr side.
\end{itemize}

The CuCr1Zr-CuCr1Zr interface was clearly discernible, homogeneous and without significant voids population. For both interfaces, the interfacial grain boundaries remained in the plane of the joint, therefore the diffusion bonding cannot be considered fully complete according to the above definition. Nevertheless, the measurements of thermal conductivity and mechanical strength, for the CuCr1Zr-SS~316L interface, showed values limited by the SS~316L and comparable to its bulk whereas for the CuCr1Zr-CuCr1Zr interface the values were comparable to the bulk of precipitation hardened CuCr1Zr. For these reasons, it can be concluded that the diffusion bonding reached advanced stage (phase 2). The data obtained from the test bench showed that the manufactured real size prototype is well capable of withstanding the thermal power that the most critical of the SPS dump’s cooling plates would undergo in the worst operational scenario. Finally, the homogeneity of the cooling performance suggests that the HIP diffusion bonding process, for the corresponding interfaces all over the prototype plate, is homogeneous.

\section{Conclusions}
The future CERN SPS internal dump, TIDVG\#5, will have to cope with higher beam intensities and repetition dumping rates starting from 2021. The maximum average thermal power carried by the beam can be as high as approximately 270~kW and about 90\% of it is absorbed by the dump and shall be evacuated by the cooling plates. Their cooling performance is, therefore, crucial. Diffusion bonding of stainless steel~316L and precipitation hardened CuCr1Zr by HIPing was thoroughly investigated as manufacturing technique for the dump’s cooling plates to reduce interfacial thermal resistance between the different components.
Micro-structural analyses and measurements of thermal conductivity and mechanical strength suggest that diffusion bonding at the interfaces CuCr1Zr-SS316L and CuCr1Zr-CuCr1Zr reached an advanced stage. Moreover, a test bench allowed to measure the cooling performance of the real size prototype with the worst operational scenario and proved that the cooling performance are uniform and the manufacturing technique can be used for the real cooling plates of the future SPS internal dump.
In conclusion, the present work opens up the way for further investigations on the exploitability of this technique for other challenging cooling applications in the field of high energy physics.
\pagebreak
\bibliography{TIDVG_HIP}

\end{document}